\documentclass[reprint,aps,prl,superscriptaddress,nofootinbib,floatfix]{revtex4-2}

\makeatletter\@booleantrue\acknowledgments@sw\makeatother

\usepackage{amsmath,amssymb,mathtools}
\usepackage{graphicx}
\usepackage{tikz}
\usepackage{xcolor}
\usepackage{hyperref}
\hypersetup{colorlinks=true,linkcolor=blue!60!black,citecolor=blue!60!black,urlcolor=blue!60!black}

\usepackage{amsthm}
\theoremstyle{definition}
\newtheorem{theorem}{Theorem}
\newtheorem*{lemma}{Lemma}
\newtheorem*{corollary}{Corollary}

\newcommand{\Tr}{\operatorname{Tr}}
\newcommand{\PhiP}{\Phi^{+}}
\newcommand{\Xtwirl}{\mathcal X}

\newcommand{\nngWidthMaxExact}{0.0283}

\newcommand{\nngLamMinusPi}{0.6011}

\definecolor{RED}{rgb}{1,0,0}
\begin{document}

\title{One Relative Phase Orders Operational Thresholds of Noisy Bell Pairs}

\author{Xuan Du Trinh}
\email{xtrinh@cs.stonybrook.edu}
\affiliation{Stony Brook University, Stony Brook, New York 11794, USA}
\date{\today}

\begin{abstract}
Unlike the two-qubit entangled and CHSH-nonlocal sets, the steerable and
Bell-nonlocal ones have no known general closed-form criterion.  For any such operational ability whose failing
states form a convex set closed under local unitaries,
and for a Bell mixture with arbitrary noise, the $X$ part of the noise gives an upper bound on the
definitive threshold, the Bell weight above which the mixture attains the
ability.  That bound is the tightest that the seven of the fifteen Pauli
expectation values fixing the $X$ part can support.  The same theorem shows
that a state has the ability whenever its $X$ part does, so partial
tomography of a state can certify its ability.  When noise is of $X$ form, the mixture carries a relative phase between the noise and
Bell coherences that cannot be gauged away by local unitaries, and it controls
the threshold through interference.  We prove a monotonicity law: the threshold is
nondecreasing in this phase, whether or not a closed-form
criterion for the ability exists.
\end{abstract}

\maketitle

Maximally entangled states are the elementary resource for quantum
communication~\cite{BennettBrassardMermin1992,CurtyLewensteinLutkenhaus2004,
ZukowskiEtAl1993}.  However, a Bell state is rarely distributed in perfect
condition because of decoherence.  As the noise level increases, the
operational abilities of the distributed state are lost in stages.  A Bell
state enables unit-fidelity teleportation, is steerable,
and attains the maximal CHSH violation, whereas a noisy state may lose these
abilities at different noise levels.  Teleportation usefulness and
steerability are distinct operational abilities, and each can occur without
the other~\cite{OperationalLayersCompanion}.

Steerability rules out all local-hidden-state models for the conditional
states prepared on qubit $B$ by measurements on qubit $A$ in a
two-qubit system~\cite{WisemanJonesDoherty2007}.  Bell nonlocality means that there exist
local measurements on the state whose joint statistics cannot be
explained by any local-hidden-variable model~\cite{BrunnerEtAl2014}.  These two abilities license
different levels of hardware trust, from one-sided device-independent to fully
device-independent key distribution~\cite{BranciardEtAl2012,AcinEtAl2007},
and both regimes have been demonstrated~\cite{WalkEtAl2016,SaundersEtAl2010,
HandchenEtAl2012,HensenEtAl2015,ShalmEtAl2015,NadlingerEtAl2022}.
Entanglement and teleportation usefulness have closed-form criteria for
two-qubit states~\cite{Peres1996,Horodecki1996,HorodeckiTeleportation1999}.  By
contrast, no general closed-form criterion decides whether an arbitrary
two-qubit state is steerable or Bell nonlocal~\cite{WisemanJonesDoherty2007,JonesWisemanDoherty2007,
BrunnerEtAl2014}.

\begin{figure}[t]
\centering
\includegraphics[width=\columnwidth]{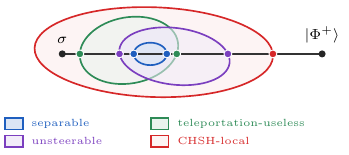}
\caption{Schematic cross-section of ability-absence sets for two-qubit states.
CHSH-local means that the CHSH value, optimized over local measurement
settings, does not exceed the local bound $2$.  Separable states lie in every
depicted set.  The teleportation-useless and unsteerable sets lie inside the
CHSH-local set~\cite{HorodeckiTeleportationBell1996,WisemanJonesDoherty2007}.  The
teleportation-useless and unsteerable sets overlap without either containing
the other~\cite{OperationalLayersCompanion}.  The segment is the
Bell-mixing line of Eq.~\eqref{eq:line} from $\PhiP$ at $\lambda=1$
to $\sigma$ at $\lambda=0$, and each colored point marks an edge of the
corresponding ability-absence interval.  The figure illustrates
an example in which $\sigma$ is steerable and teleportation-useful but CHSH-local.
Thus, the CHSH-local interval contains $\lambda=0$.  The separable,
teleportation-useless, and unsteerable intervals do not.  They describe loss and
recovery of entanglement, teleportation usefulness, and steerability along the
segment~\cite{OperationalLayersCompanion}.}
\label{fig:hierarchy}
\end{figure}

To analyze how noisy states cross these boundaries as the noise fraction
increases, whether or not closed-form criteria are available, we consider
the Bell-mixing line
\begin{equation}
\rho_\lambda(\sigma)=\lambda\PhiP
+(1-\lambda)\sigma,
\qquad 0\le\lambda\le1,
\label{eq:line}
\end{equation}
where $\lambda$ is the Bell weight, $\sigma$ is the noise state, and
$\PhiP=|\PhiP\rangle\!\langle\PhiP|$ is the Bell-state density operator, where
$|\PhiP\rangle=(|00\rangle+|11\rangle)/\sqrt2$.  Increasing
the noise fraction, i.e., decreasing $\lambda$ from $1$ to $0$, moves the state
from $\PhiP$ to $\sigma$.  Distinct abilities can be lost at
different Bell weights, and an ability can be lost and then recovered along
the line~\cite{OperationalLayersCompanion} (Fig.~\ref{fig:hierarchy}).
At a Bell weight, the practical question is
therefore which operational boundaries the noisy mixture has crossed, and how
those boundaries move when the noise state varies.
This problem is unavoidable, because operational crossings decide
whether a quantum communication protocol provides an advantage.

The archetype is the Werner family with white noise $\sigma=I/4$.  It has a hierarchy of thresholds, the Bell
weights above which each ability holds: $\lambda=1/3$ for
entanglement and teleportation usefulness, $1/2$ for
steerability, $1/\sqrt3$ for violation of the optimized three-setting CJWR
inequality, and $1/\sqrt2$ for violation of the optimized CHSH
inequality~\cite{Werner1989,
HorodeckiTeleportationBell1996,WisemanJonesDoherty2007,
ZhangChitambar2024,CavalcantiJonesWisemanReid2009,CostaAngelo2016,
HorodeckiCHSH1995}.
Settling this short list for a one-parameter family took thirty-five
years, from Werner's construction to the recent proof that the steerability
threshold is not smaller for positive operator-valued measures (POVMs) than
for projective measurements~\cite{Werner1989,ZhangChitambar2024}.
Outside the Werner family, locating each such boundary generally requires
all fifteen parameters of the noise state, and even then these
boundaries are largely unknown.

We address both obstacles by considering the boundaries when
replacing the noise state $\sigma$ by its $X$ part in Eq.~\eqref{eq:line}.
First, we prove that the $X$ part attains the largest operational
threshold among all noise states sharing that $X$ part (Theorem~\ref{thm:shadow}). For
general two-qubit states, knowing the $X$ part alone gives the optimal uniform
guarantee: if the $X$ part has an ability, every compatible noise state
inherits it.
Second, $\PhiP$ is itself of $X$ form, so the replacement leaves the
whole mixture in $X$ form, and various analytic thresholds and criteria are known for $X$
states~\cite{Hu2013XStates,eac-paper,OperationalLayersCompanion}.  Thus,
accepting the upper bound rather than demanding the exact threshold gives
access to a rich class of computable boundaries.
Equation~\eqref{eq:line} describes a noise-exposure process in which $\PhiP$
remains undisturbed with probability $\lambda$, while the exposure decoheres
it to $\sigma$ with probability $1-\lambda$. Standard local amplitude damping,
pure dephasing, and their combination all preserve the $X$ form during the
evolution from $\PhiP$ to $\sigma$~\cite{NielsenChuang2010,YuEberly2004,YuEberly2007}.  When the noise
mechanism is not $X$-form preserving, we also quantify the resulting precision
loss from using the $X$ part.

We go further in characterizing the movement of the boundaries when the
seven parameters of the $X$ noise state vary.  We obtain a monotonicity
law (Theorem~\ref{thm:phase}) in the relative phase $\theta$ between the Bell
coherence and the noise coherence
$u=\langle00|\sigma|11\rangle=|u|e^{i\theta}$, and surprisingly this law holds for every
operational ability whose failing states form a convex set closed under local
unitaries, regardless of whether a closed-form criterion exists.
At fixed populations and coherence moduli, any such threshold does not
decrease with $\theta$, so it is smallest for aligned noise ($\theta=0$) and
largest for anti-aligned noise ($\theta=\pi$).
The $X$-part upper bound for arbitrary noise is therefore itself
nondecreasing in $\theta$.

\paragraph*{Setup.}
We use \emph{ability} as a common label for properties such as
entanglement, steerability,
Bell nonlocality, optimized witness violation, and positivity of a protocol
lower bound.  We state the optimization, direction, and measurement class
where applicable.  For a chosen ability, let $K$ be the ability-absence set, the set
of states that lack it.  The results below cover each such $K$ that is convex and closed under
local unitaries.  We define the corresponding ability-absence
interval and \emph{definitive threshold} by
\begin{align}
I_K(\sigma)
&=\{\lambda\in[0,1]:\rho_\lambda(\sigma)\in K\},\notag\\
\lambda_K(\sigma)
&=\sup I_K(\sigma),\qquad \sup\emptyset:=0 .
\label{eq:threshold}
\end{align}

Convexity of $K$ and affinity of Eq.~\eqref{eq:line} make $I_K(\sigma)$ an
interval, possibly empty and not necessarily containing $\lambda=0$, with
$\lambda_K=0$ when it is empty.  Every $K$
considered here is a closed set (SM, Sec.~S\ref{sec:sm-phase-proof}), so if $\sigma\in K$, then $I_K(\sigma)=[0,\lambda_K]$
and the ability is absent through $\lambda_K$ and present above it, as in
the Werner family.

A two-qubit $X$ state $\sigma_X$ has nonzero entries only on the diagonal
and the anti-diagonal in the computational basis as follows
\begin{equation}
\setlength{\arraycolsep}{8pt}
\renewcommand{\arraystretch}{1.15}
\sigma_X=
\begin{pmatrix}
a&0&0&u\\
0&b&v&0\\
0&\bar v&c&0\\
\bar u&0&0&d
\end{pmatrix},
\quad
\begin{gathered}
a,b,c,d\ge0,\\[2pt]
a+b+c+d=1,\\[2pt]
ad\ge|u|^{2},\\[2pt]
bc\ge|v|^{2},
\end{gathered}
\label{eq:xstate}
\end{equation}
where $u$ and $v$ are the coherences of the even-parity sector
$\{|00\rangle,|11\rangle\}$ and the odd-parity sector
$\{|01\rangle,|10\rangle\}$.

\paragraph*{The $X$ part and its induced threshold upper bound.}

Define the $X$-twirl map
\begin{equation}
\Xtwirl:\ \tau\mapsto
\frac12\left[\tau+
(\sigma_z\otimes\sigma_z)\tau(\sigma_z\otimes\sigma_z)\right].
\label{eq:xtwirl}
\end{equation}
The map is a Pauli channel that deletes the coherences outside the
anti-diagonal of the input.  Its fixed points include $\PhiP$, and it
commutes with the mixing map
$\tau\mapsto\rho_\lambda(\tau)$ of Eq.~\eqref{eq:line}.  For a
physical noise state $\sigma$, the $X$-twirl returns the $X$ part
$\sigma_X=\Xtwirl(\sigma)$ of Eq.~\eqref{eq:xstate}, which is
reconstructed from the seven Pauli expectation values of $\sigma$,
$\langle\sigma_z\otimes I\rangle$, $\langle I\otimes\sigma_z\rangle$,
$\langle\sigma_z\otimes\sigma_z\rangle$,
$\langle\sigma_x\otimes\sigma_x\rangle$,
$\langle\sigma_x\otimes\sigma_y\rangle$,
$\langle\sigma_y\otimes\sigma_x\rangle$, and
$\langle\sigma_y\otimes\sigma_y\rangle$.  However, different states can share
the same $X$ part.  Measuring them takes five of the nine local Pauli measurement
settings that full tomography uses.  The same seven observables read on the delivered mixture give
$\rho_\lambda(\sigma_X)$.

\begin{theorem}[$X$-part minimax]\label{thm:shadow}

Denote the set of noise states that share the same
$X$ part $\sigma_X$ by
$\Omega(\sigma_X)=\{\tau:\Xtwirl(\tau)=\sigma_X\}$, where $\tau$ ranges over
physical two-qubit states.  For every convex
ability-absence set $K$ closed under local unitaries and every
$\tau\in\Omega(\sigma_X)$,
\begin{equation}
I_K(\tau)\subseteq I_K(\sigma_X),
\qquad
\lambda_K(\tau)\le\lambda_K(\sigma_X).
\label{eq:shadow-inclusion}
\end{equation}
Moreover, $\max_{\tau\in\Omega(\sigma_X)}\lambda_K(\tau)
=\lambda_K(\sigma_X)$.
\end{theorem}

\textit{Proof.} Fix $\tau\in\Omega(\sigma_X)$ and $\lambda\in[0,1]$.
Linearity and $\Xtwirl(\PhiP)=\PhiP$ give
$\Xtwirl\bigl(\rho_\lambda(\tau)\bigr)
=\rho_\lambda\bigl(\Xtwirl(\tau)\bigr)
=\rho_\lambda(\sigma_X)$.
If $\rho_\lambda(\tau)\in K$, local-unitary closure puts its conjugate
by $\sigma_z\otimes\sigma_z$ in $K$, and convexity puts their equal mixture
$\Xtwirl(\rho_\lambda(\tau))$ in $K$.  The identity above then gives
$\rho_\lambda(\sigma_X)\in K$.  Thus,
$I_K(\tau)\subseteq I_K(\sigma_X)$, and taking suprema gives the threshold
inequality.  Finally, $\sigma_X\in\Omega(\sigma_X)$, so the upper bound is attained by
$\sigma_X$ itself.~$\square$

At a fixed Bell weight $\lambda$,
$\{\rho_\lambda(\tau):\tau\in\Omega(\sigma_X)\}\cap K=\emptyset$ if and only if
$\rho_\lambda(\sigma_X)\notin K$,
since Theorem~\ref{thm:shadow} transfers membership in $K$ and
$\sigma_X\in\Omega(\sigma_X)$.  Hence, if $\lambda>\lambda_K(\sigma_X)$, this
intersection is empty, and $\lambda_K(\sigma_X)$ is the tightest upper bound on
the definitive threshold $\lambda_K(\sigma)$ that the seven expectation values
of $\sigma$ support (SM, Sec.~S\ref{sec:sm-shadow}).
At $\lambda=0$, Eq.~\eqref{eq:shadow-inclusion} gives the direct
inheritance rule $\Xtwirl(\tau)=\sigma_X\notin K\Rightarrow\tau\notin K$.
The rule implies that any state, not just the noise state, has the
ability whenever its $X$ part does.  Thus, partial tomography of a state
can certify its operational ability.
We now characterize the cost in precision of computing $I_K$ and $\lambda_K$
from $\sigma_X$ rather than from the full noise state $\sigma$, namely the
threshold gap $\delta_K(\sigma):=\lambda_K(\sigma_X)-\lambda_K(\sigma)$
and the measure
$\mu_K(\sigma):=\operatorname{meas}[I_K(\sigma_X)\setminus I_K(\sigma)]$.
Both are nonnegative by Theorem~\ref{thm:shadow} and both vanish on the
$X$ family.
Write $\varepsilon=\tfrac12\|\sigma-\sigma_X\|_1$ and let
$\mathcal D_K(\sigma_X)=\inf_{\omega\notin K}\tfrac12\|\sigma_X-\omega\|_1$ be
the distance from $\sigma_X$ to the nearest unit-trace Hermitian operator,
not necessarily a state, outside $K$.  If
$\mathcal D_K(\sigma_X)>0$, so that $\sigma_X$ lies strictly inside $K$, then
$\max\{\delta_K(\sigma),\mu_K(\sigma)\}
\le\varepsilon\,\lambda_K(\sigma_X)/\mathcal D_K(\sigma_X)$.
The bound is linear in $\varepsilon$ and conservative.  The analysis, including the case
$\sigma_X\notin K$ in which the ability is already present at $\lambda=0$ and
can be lost and later regained, is given in the SM
(Sec.~S\ref{sec:sm-tightness}).

\paragraph*{One relative coherence phase.}
We now turn to the second question: how the boundaries move as the
parameters of the $X$ noise state vary.  We first consider the coherences
$u$ and $v$.  In the literature on quantum
discord~\cite{AliRauAlber2010,Chen2011XDiscord} and entanglement dynamics of
$X$ states~\cite{QuesadaAlQasimiJames2012,YuEberly2007}, their phases are
degrees of freedom that local unitaries gauge away while closed-form results
are still obtained.  Conjugation by
$W(\alpha,\beta)=\mathrm{diag}(1,e^{i\alpha})\otimes\mathrm{diag}(1,e^{i\beta})$,
which we call the local phase gauge, multiplies the even-sector coherence of an $X$
state by $e^{-i(\alpha+\beta)}$ and the odd-sector one by
$e^{-i(\alpha-\beta)}$.  At fixed $(a,b,c,d,|u|,|v|)$, the mixture
$\rho_\lambda(\sigma_X)$ is itself an $X$ state, with even-sector coherence
$z_\lambda(u)=\tfrac{\lambda}{2}+(1-\lambda)u$ and odd-sector coherence
$(1-\lambda)v$, so
$W\bigl(\tfrac{\arg z_\lambda+\arg v}{2},\tfrac{\arg z_\lambda-\arg v}{2}\bigr)$
makes both of them real and nonnegative.  This removes $\arg v$ from the phase
dependence of the threshold problem, but the relative phase $\theta=\arg u$
survives in the modulus $q_\lambda(\theta)=|z_\lambda(u)|$.  We call $\theta$ a relative phase because every gauge
rotation turns $u$ and the Bell coherence
$\langle00|\PhiP|11\rangle=\tfrac12$ by the same angle.  For $0<\lambda<1$
and $|u|>0$, $\cos\theta$ is a local-unitary invariant
(SM, Sec.~S\ref{sec:sm-phase-proof}).

Because $q_\lambda(\theta)^{2}=\tfrac{\lambda^{2}}{4}
+\lambda(1-\lambda)|u|\cos\theta+(1-\lambda)^{2}|u|^{2}$ depends on $\theta$ only through $\cos\theta$, the states at
$\pm\theta$ are locally unitarily equivalent at each Bell weight, and throughout we take
$\theta\in[0,\pi]$, where $q_\lambda(\theta')\le q_\lambda(\theta)$ for
$0\le\theta\le\theta'\le\pi$.  The following lemma realizes this scalar decrease as a
mixture of local-unitary images.

\begin{lemma}[even-sector coherence reduction]
Let $\rho$ and $\rho'$ be two-qubit $X$ states with the same
populations, the same odd-sector coherence modulus, and even-sector
coherence moduli $q\ge q'$.  Then $\rho'$ is a convex
mixture of local-unitary images of $\rho$.
\end{lemma}

The proof comes from an identity: once the two states are in the normal form with
real nonnegative coherences,
$\rho'=\tfrac12\,W_\varphi\,\rho\,W_\varphi^\dagger
+\tfrac12\,W_{-\varphi}\,\rho\,W_{-\varphi}^\dagger$
with $W_\varphi=W(\varphi,\varphi)$ and $\cos2\varphi=q'/q$ when $q>0$.
If $q=q'=0$, then $\rho'$ is a local-unitary image of $\rho$.  Let $\sigma_X(\theta)$ be the noise state at
fixed $(a,b,c,d,|u|,|v|)$ that carries $u=|u|e^{i\theta}$.  At each fixed
$\lambda$, the lemma therefore writes $\rho_\lambda(\sigma_X(\theta'))$ as
a convex mixture of local-unitary images of
$\rho_\lambda(\sigma_X(\theta))$.
The conversion runs one way, producing the larger $\theta'$ from the smaller
$\theta$ with a coin shared between the two parties $A$ and $B$
(SM, Sec.~S\ref{sec:sm-contraction}).  That conversion is the key
observation behind the phase dependence of the ability boundaries.

With convexity
and local-unitary closure of $K$, a weight $\lambda$ that puts
$\rho_\lambda(\sigma_X(\theta))$ in $K$ also puts the mixture
$\rho_\lambda(\sigma_X(\theta'))$ in $K$, so
$I_K(\sigma_X(\theta))\subseteq I_K(\sigma_X(\theta'))$ and the definitive
threshold, the supremum of the interval, is nondecreasing in $\theta$
(SM, Sec.~S\ref{sec:sm-phase-proof}).  Reference~\cite{OperationalLayersCompanion} observed
this monotonicity by direct computation with the closed-form intervals, such as the
separable and the CHSH-local ones.  Our reasoning here involves no such formula,
so the theorem below covers the unsteerable and Bell-local intervals as well,
where no closed-form criterion is known.

\begin{theorem}[relative-phase monotonicity law]\label{thm:phase}
Let $K$ be any convex ability-absence set closed under local
unitaries, and let $\sigma_X(\theta)$ carry $u=|u|e^{i\theta}$ at fixed
populations $(a,b,c,d)$ and coherence moduli $(|u|,|v|)$. For
$0\le\theta\le\theta'\le\pi$,
\begin{align}
I_K(\sigma_X(\theta))&\subseteq I_K(\sigma_X(\theta')),\notag\\
\lambda_K(\sigma_X(\theta))&\le\lambda_K(\sigma_X(\theta')).
\label{eq:phase-order}
\end{align}
\end{theorem}

Theorem~\ref{thm:phase} does not require the noise to lack the chosen
ability. If the noise has it, the ability-absence interval may be empty, or
detached from $\lambda=0$ with a loss and later revival
(Fig.~\ref{fig:hierarchy}). The theorem describes the widening of that whole
interval with $\theta$.

\paragraph*{Operational consequences.}

The abilities whose ability-absence sets are convex and local-unitary
closed include entanglement, standard teleportation
usefulness~\cite{HorodeckiTeleportationBell1996}, steerability in either
direction and Bell nonlocality with projective measurements or
POVMs~\cite{WisemanJonesDoherty2007,
JonesWisemanDoherty2007,BrunnerEtAl2014,Barrett2002,QuintinoEtAl2015,
ZhangChitambar2024}, optimized three-setting CJWR violation, and optimized
CHSH violation~\cite{CavalcantiJonesWisemanReid2009,CostaAngelo2016,
HorodeckiCHSH1995} (SM, Sec.~S\ref{sec:sm-phase-proof}).
Bell nonlocality and steerability are state properties, whereas the
optimized CJWR and CHSH violations are specific witnesses for them.  The two violations are
computable, and if they do not occur, the properties can still be present.
The list is not
exhaustive: the proof of Theorem~\ref{thm:phase} uses only convexity and
local-unitary closure of the ability-absence set, so any ability formulated
in the future with these two properties inherits the same monotonicity law,
whether or not it has a closed-form criterion.

\begin{figure}[t]
\centering
\includegraphics[width=\columnwidth]{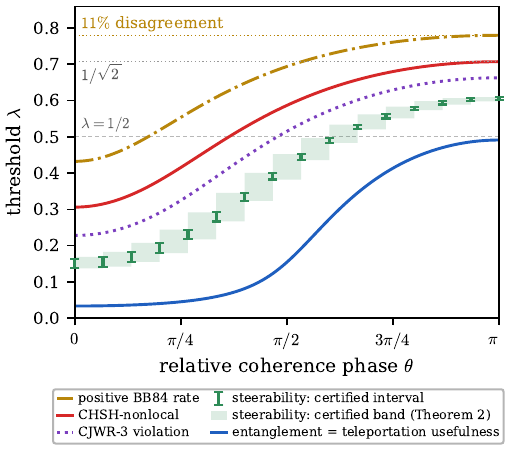}
\caption{The operational thresholds are nondecreasing in the relative
coherence phase $\theta$. Analytic curves give the entanglement and coincident
teleportation usefulness thresholds, the three-setting CJWR witness
threshold, the CHSH-nonlocal threshold, and the threshold for a positive key rate in
one-way BB84 with CSS codes. The last three are each optimized over the
local measurements. At $\lambda=\tfrac12$, the aligned state yields a
positive key rate, whereas the
anti-aligned state is certified $A\to B$ unsteerable for all projective
measurements. The vertical bars are certified intervals for the $A\to B$ projective-measurement
steerability threshold at their phases, from the critical-radius criterion, while the
shaded band is the region Theorem~\ref{thm:phase} certifies for phases between them
(SM, Sec.~S\ref{sec:sm-numerics}). Horizontal
guides mark equal mixing $\lambda=\tfrac12$, the weight $1/\sqrt2$ at
which a Werner state becomes CHSH-nonlocal, and the weight $0.779944$ at which
a Werner state first gives a positive key rate, where the two parties'
measurement outcomes disagree at a rate of $11.0\%$. The $X$ noise family plotted here has
$(a,b,c,d)=(0.30,0.25,0.25,0.20)$, $u=|u|e^{i\theta}$, and
$|u|=|v|=0.95\min\{\sqrt{ad},\sqrt{bc}\}$, as in
Ref.~\cite{OperationalLayersCompanion}.}
\label{fig:alignment-prl}
\end{figure}

The $A\to B$ projective-measurement steerability threshold of the family
in Fig.~\ref{fig:alignment-prl} is not known analytically.  At each sampled phase point, the critical-radius criterion of Ref.~\cite{NguyenNguyenGuhne2019}, which decides steerability by an optimization rather than a formula, locates it
inside a certified interval of width at most $\nngWidthMaxExact$ (SM, Sec.~S\ref{sec:sm-numerics}).
Beyond these abilities, in the key distribution protocol BB84, the two
parties each measure
along two orthogonal local directions on many shared noisy Bell pairs, and
the outcome disagreement rates in the two bases set
the asymptotic secret-key rate they can reach with one-way classical
communication and CSS codes~\cite{BennettBrassardMermin1992,ShorPreskill2000}.  Optimized over the local
bases, this achievable rate is convex in the state, and
the states for which it is not positive form a convex, local-unitary closed
set (SM, Sec.~S\ref{sec:sm-phase-proof}).  The Bell weight above
which a mixture with $X$ noise reaches a positive key rate is therefore
nondecreasing in $\theta$.

\paragraph*{Beyond the relative phase.}
Theorem~\ref{thm:phase} extends to changes in the coherence
moduli. For
physical $X$ noise states
$\sigma_X$ and $\sigma'_X$ with the same populations, let
$u=re^{i\theta}$ and $u'=r'e^{i\theta'}$, where
$\theta,\theta'\in[0,\pi]$. If $r'\le r$,
$r'\cos\theta'\le r\cos\theta$, and $|v'|\le|v|$, then
$I_K(\sigma_X)\subseteq I_K(\sigma'_X)$ and
$\lambda_K(\sigma_X)\le\lambda_K(\sigma'_X)$ for every convex and
local-unitary closed ability-absence set $K$. At $r'=r>0$ and $|v'|=|v|$, the hypotheses become $\theta'\ge\theta$, recovering
Theorem~\ref{thm:phase}. The proof and incomparable cases are given in the SM
(Sec.~S\ref{sec:sm-contraction}).

The seven expectation values fix both coherences of the $X$ part,
including the relative phase $\theta=\arg u$. Suppose instead that only the
noise parameters $(a,b,c,d,|u|,|v|)$ are known, and let
$\Omega_{\rm mod}$ be the nonempty class of physical states whose $X$ parts
share those values.  For
$\tau\in\Omega_{\rm mod}$ with $X$ part $\tau_X$, Theorem~\ref{thm:shadow} gives
$\lambda_K(\tau)\le\lambda_K(\tau_X)$.  The Bell-preserving gauge
$W(\alpha,-\alpha)$ rotates $v$ to $|v|$ while leaving $u$ and every threshold
unchanged, so the result is $\sigma_X(\theta)$ as in
Theorem~\ref{thm:phase}.  The monotonicity law then gives
$\lambda_K(\sigma_X(\theta))\le\lambda_K(\sigma_X(\pi))$ directly.  The anti-aligned $\sigma_X(\pi)$, which carries
$v=|v|$ in that gauge and $u=-|u|$, itself belongs to $\Omega_{\rm mod}$, so
without knowing the phase, $\lambda_K(\sigma_X(\pi))$ is the tightest uniform
upper bound on the threshold for the whole class
(SM, Sec.~S\ref{sec:sm-shadow}).

\begin{corollary}[minimax over an unknown phase]
Let $K$ be any convex ability-absence set closed under local
unitaries.  Then
\begin{equation}
\max_{\tau\in\Omega_{\rm mod}}\lambda_K(\tau)
=\lambda_K(\sigma_X(\pi)).
\label{eq:phase-blind-minimax}
\end{equation}
\end{corollary}

\paragraph*{Discussion.}

Previous work characterized analytic entanglement and standard teleportation usefulness
thresholds for arbitrary product noise and separable complex $X$
noise~\cite{eac-paper}, and analytic separable, teleportation-useless,
optimized CJWR-satisfying, and CHSH-local intervals for arbitrary complex $X$
noise~\cite{OperationalLayersCompanion}.  The latter reference observed the
widening of the ability-absence intervals with the relative phase by direct
calculation when closed-form conditions are known.  Our proof
instead needs no such per-ability calculation, and covers a large class of abilities
including directional steerability and
Bell nonlocality, whose boundaries have no known formula.

At each fixed Bell weight, the two parties can produce the more
misaligned mixture, by tossing a shared coin and each applying the phase gate it
selects, so alignment minimizes every definitive threshold considered here. The $X$-part minimax theorem turns an exact
threshold or a certified upper bound for a Bell mixture with $X$ noise into a
uniform upper bound over every compatible noise state, opening the
partial-tomography approach for operational analysis. Direct application to the Werner
family, where $\sigma=I/4$, guarantees POVM steerability in each direction above
$\lambda=1/2$ and CHSH violation above $1/\sqrt2$ for every
$\tau\in\Omega(I/4)$. The transferred bound can be exact.  For the two diagonal $X$ noise families of
Ref.~\cite{OperationalLayersCompanion} whose directional steerability
thresholds are zero, it certifies the same thresholds for every compatible noise state
(Direct Applications).  Finally, the message is not about a new formula or a single value for any
operational threshold, but about comparing and bounding thresholds without
closed-form criteria.

\newpage
\begin{acknowledgments}
The author used Claude Opus 5.0 (Anthropic) for the code behind the
steering computation of the Supplemental Material, directing each step and
checking the output.
\end{acknowledgments}

\bibliographystyle{apsrev4-2}
\bibliography{refs}

\section*{Direct Applications}

\paragraph*{Result transfer over the compatibility class.}
For any convex, local-unitary closed ability-absence set $K$, Theorem~\ref{thm:shadow}
transfers each exact threshold or certified upper bound for a physical $X$
representative into a uniform upper bound over its complete compatibility
class $\Omega(\sigma_X)$. Specifically, for every physical $X$ state $\sigma_X$ and every
$\tau\in\Omega(\sigma_X)$,
\begin{equation}
0\leq\lambda_K(\tau)\leq\lambda_K(\sigma_X).
\label{eq:end-fiber-transfer}
\end{equation}
Thus, an ability of the representative mixture at a fixed Bell weight is
inherited by every compatible mixture. If
$\lambda_K(\sigma_X)=0$, Eq.~\eqref{eq:end-fiber-transfer} forces
$\lambda_K(\tau)=0$ throughout the
class, so the ability is present at every positive Bell weight even when it is
absent at $\lambda=0$. If, however, $\sigma_X$ has the ability at
$\lambda=0$, then so does every compatible noise state $\tau$, including the
original $\sigma$.

\paragraph*{The Werner class.}
The $X$ representative $\sigma_X=I/4$ corresponds to the Werner family
$\rho_\lambda(I/4)$. The complete compatibility class $\Omega(I/4)$ consists
of all noise states of the computational-basis form
\begin{equation}
\tau=
\begin{pmatrix}
\tfrac14&x&y&0\\
x^*&\tfrac14&0&z^*\\
y^*&0&\tfrac14&w^*\\
0&z&w&\tfrac14
\end{pmatrix}.
\label{eq:end-werner-fiber}
\end{equation}
Reordering to the parity basis
$(|00\rangle,|11\rangle,|01\rangle,|10\rangle)$ collects the four free entries
into an off-diagonal block,
$$
[\tau]_{\rm par}=
\begin{pmatrix}
I_2/4&C\\
C^\dagger&I_2/4
\end{pmatrix},
\qquad
C=\begin{pmatrix}x&y\\ z&w\end{pmatrix},
$$
and $\tau$ is a state exactly when the singular values of $C$ are at most
$\tfrac14$.  Written out in the entries themselves, with
$s=|x|^{2}+|y|^{2}+|z|^{2}+|w|^{2}$, that condition is
$$
s+\sqrt{s^{2}-4\,|xw-yz|^{2}}\ \leq\ \tfrac18 .
$$
In this class, the seven Pauli expectation values fixing the $X$ part are
all zero. Subject only to the positivity condition above, the four complex
parameters $x,y,z,w$ describe a large family of compatible states.
For the Werner family, the exact steerability threshold is $1/2$ in both
directions, both for projective measurements~\cite{WisemanJonesDoherty2007}
and for POVMs~\cite{ZhangChitambar2024}, and the CHSH-nonlocal threshold is
$1/\sqrt2$~\cite{HorodeckiCHSH1995}. Equation~\eqref{eq:end-fiber-transfer}
therefore gives, for every state in Eq.~\eqref{eq:end-werner-fiber},
\begin{equation}
\begin{aligned}
\lambda>\frac12
&\ \Longrightarrow\ \rho_\lambda(\tau)
\text{ is two-way POVM-steerable},\\
\lambda>\frac1{\sqrt2}
&\ \Longrightarrow\ \rho_\lambda(\tau)
\text{ violates CHSH}.
\end{aligned}
\label{eq:end-werner-guarantees}
\end{equation}
\paragraph*{Beyond the Werner class.}
Reference~\cite{OperationalLayersCompanion} proves that both directional
steerability thresholds, for projective measurements or POVMs,
are zero for the two diagonal families
\begin{equation}
\sigma_X=\operatorname{diag}(a,b,0,d),
\qquad
\sigma_X=\operatorname{diag}(a,0,c,d),
\label{eq:end-zero-families}
\end{equation}
where the entries are nonnegative and sum to one. Thus, every state in the
corresponding compatibility classes has zero
steerability thresholds. Here, the transferred value is zero, so
the two bounds in Eq.~\eqref{eq:end-fiber-transfer} coincide and the seven Pauli expectation
values fix these thresholds exactly rather than bounding them. In the first class, positivity forces the row and
column associated with the zero population to vanish. The general compatible
state therefore has the form
\begin{equation}
\tau=
\begin{pmatrix}
a&x&0&0\\
x^*&b&0&y\\
0&0&0&0\\
0&y^*&0&d
\end{pmatrix}\geq0,
\qquad a+b+d=1,
\label{eq:end-nonx-zero-family}
\end{equation}
where $x$ and $y$ are the remaining coherences, and positivity is
equivalent to $|x|^2\leq ab$, $|y|^2\leq bd$, and
$d|x|^2+a|y|^2\leq abd$. These states are generally non-$X$.

\clearpage
\onecolumngrid

\begin{center}
\textbf{\large Supplemental Material for\\
``One Relative Phase Orders Operational Thresholds of Noisy Bell Pairs''}
\\[4pt]
Xuan Du Trinh\\
\textit{Stony Brook University, Stony Brook, New York 11794, USA}
\end{center}

\setcounter{figure}{0}
\renewcommand{\thefigure}{S\arabic{figure}}
\renewcommand{\theHfigure}{S\arabic{figure}}
\setcounter{equation}{0}
\renewcommand{\theequation}{S\arabic{equation}}
\renewcommand{\theHequation}{S\arabic{equation}}
\setcounter{table}{0}
\renewcommand{\thetable}{S\arabic{table}}
\renewcommand{\theHtable}{S\arabic{table}}
\newcounter{smsection}

\vspace{0.5em}

This Supplemental Material supports the main text as follows.
Section~S\ref{sec:sm-shadow} proves Theorem~\ref{thm:shadow} and gives the
formulas that reconstruct the $X$ part from the seven Pauli expectation
values.  It then proves that $\lambda_K(\sigma_X)$ is the smallest uniform upper
bound those values support, and the corollary that the anti-aligned
threshold $\lambda_K(\sigma_X(\pi))$ is the tightest upper bound when the
coherence phase is unknown.  Section~S\ref{sec:sm-tightness} proves the precision-loss
bound quoted in the main text, and covers the case $\sigma_X\notin K$, in
which the ability is present at $\lambda=0$ and can be lost and regained.
Section~S\ref{sec:sm-contraction} proves the coherence-reduction lemma and gives
the channel that produces the more misaligned mixture at fixed Bell
weight.  The parties realize that channel with the shared coin.  The section then
proves the extension to the case where the coherence moduli are not fixed,
with the incomparable cases mentioned.
Section~S\ref{sec:sm-phase-proof} proves the local-unitary equivalence
behind the restriction to $\theta\in[0,\pi]$, and that no local unitary
removes the relative phase.  For each ability named in the main text, it
verifies convexity, local-unitary closure, and closedness of the
ability-absence set.  Closedness is what makes $I_K(\sigma)$ a closed
interval.  Those abilities include positivity of the measurement-basis-optimized asymptotic lower bound for one-way BB84 with CSS codes.
Section~S\ref{sec:sm-numerics} gives the
analytic curves and the certified $A\to B$ steerability intervals of
Fig.~\ref{fig:alignment-prl}.

\refstepcounter{smsection}
\subsection*{S\arabic{smsection}. Theorem~\ref{thm:shadow} and minimax optimality}
\label{sec:sm-shadow}

\subsubsection*{The \texorpdfstring{$X$}{X}-twirl and Pauli reconstruction}

The operator $\sigma_z\otimes\sigma_z$ has eigenvalue $+1$ on
$\mathrm{span}\{|00\rangle,|11\rangle\}$ and $-1$ on
$\mathrm{span}\{|01\rangle,|10\rangle\}$.  Therefore, the map
$\Xtwirl(\tau)=\tfrac12[\tau+(\sigma_z\otimes\sigma_z)\tau
(\sigma_z\otimes\sigma_z)]$ deletes exactly the matrix elements
connecting the two parity sectors and reproduces the $X$ entries.  The map
is an equal mixture of two local-unitary conjugations.
Because $(\sigma_z\otimes\sigma_z)\PhiP
(\sigma_z\otimes\sigma_z)=\PhiP$,
$\Xtwirl$ fixes the Bell projector, and linearity ensures that
$\Xtwirl$ commutes with the mixing map.  These two facts give the
identity $\Xtwirl(\rho_\lambda(\tau))=\rho_\lambda(\sigma_X)$.  The proof of
Theorem~\ref{thm:shadow} combines it with convexity and local-unitary closure
of $K$ to transfer membership in $K$ from $\rho_\lambda(\tau)$ to
$\rho_\lambda(\sigma_X)$.

Let
\begin{equation}
A_z=\langle\sigma_z\otimes I\rangle,\qquad
B_z=\langle I\otimes\sigma_z\rangle,\qquad
T_{ij}=\langle\sigma_i\otimes\sigma_j\rangle .
\label{eq:pauli-z-data}
\end{equation}
The seven expectation values, with normalization, fix the following
positions in the Pauli-product table of every $\tau\in\Omega(\sigma_X)$:
$$
\begin{array}{c|cccc}
\langle P_A\otimes P_B\rangle_{\tau}
& I & \sigma_x & \sigma_y & \sigma_z\\ \hline
I        & 1   & \ast   & \ast   & B_z\\
\sigma_x & \ast & T_{xx} & T_{xy} & \ast\\
\sigma_y & \ast & T_{yx} & T_{yy} & \ast\\
\sigma_z & A_z & \ast   & \ast   & T_{zz}
\end{array} .
$$
The $X$-twirl sets the eight $\ast$ entries to zero, since each is the
expectation value of a Pauli product that anticommutes with
$\sigma_z\otimes\sigma_z$.  For a compatible state, they are constrained only by
positivity of the full density matrix, which defines $\Omega(\sigma_X)$.
The populations are
\begin{equation}
\begin{aligned}
a&=\frac{1+A_z+B_z+T_{zz}}{4}, &
b&=\frac{1+A_z-B_z-T_{zz}}{4},\\
c&=\frac{1-A_z+B_z-T_{zz}}{4}, &
d&=\frac{1-A_z-B_z+T_{zz}}{4}.
\end{aligned}
\label{eq:population-reconstruction}
\end{equation}
The coherences are reconstructed from
\begin{equation}
\begin{aligned}
\operatorname{Re}u&=\frac{T_{xx}-T_{yy}}{4}, &
\operatorname{Re}v&=\frac{T_{xx}+T_{yy}}{4},\\
\operatorname{Im}u&=-\frac{T_{xy}+T_{yx}}{4}, &
\operatorname{Im}v&=\frac{T_{xy}-T_{yx}}{4}.
\end{aligned}
\label{eq:coherence-reconstruction}
\end{equation}
The correlation tensor $T$ determines the optimized CHSH and CJWR witnesses,
as well as the standard teleportation usefulness.  Exact directional
steerability depends in addition on the local marginal expectation
values, which for $X$ states are $A_z$ and $B_z$.  This dependence is visible in the criterion used in
Sec.~S\ref{sec:sm-numerics}, whose ratio depends on the Bloch vector of
$A$ as well as on $T$.

Theorem~\ref{thm:shadow} takes the seven expectation values of the noise
state $\sigma$ as input.  If instead they are known for the delivered mixture
$\rho_\lambda(\sigma)$, then Eq.~\eqref{eq:line} with known $\lambda<1$ gives for
each of the seven Pauli products $P$,
\begin{equation}
\langle P\rangle_\sigma
=\frac{\langle P\rangle_{\rho_\lambda(\sigma)}
-\lambda\langle P\rangle_{\PhiP}}{1-\lambda}.
\label{eq:delivered-inversion}
\end{equation}
Certifying the delivered mixture itself, through the inheritance rule of
the main text, needs no value of $\lambda$, whereas recovering $\sigma_X$, and
with it the endpoint threshold $\lambda_K(\sigma_X)$, needs $\lambda$ through
Eq.~\eqref{eq:delivered-inversion}.

\subsubsection*{Minimax optimality}

Let $\mathcal B(\sigma_X)$ be any threshold upper
bound that depends only on the seven expectation values defining $\sigma_X$ and
is valid for every $\tau\in\Omega(\sigma_X)$.  Being an upper bound for every
compatible state gives the inequality, and Theorem~\ref{thm:shadow} gives the
equality in
$$
\mathcal B(\sigma_X)\geq
\max_{\tau\in\Omega(\sigma_X)}\lambda_K(\tau)
=\lambda_K(\sigma_X),
$$
so $\lambda_K(\sigma_X)$ is the smallest such uniform upper bound.  For an
individual compatible state, $\lambda_K(\tau)\leq\lambda_K(\sigma_X)$ can be strict because the
interparity coherences removed by $\Xtwirl$ may contribute to the ability.
Section~S\ref{sec:sm-tightness} quantifies this gap.

The class $\Omega(\sigma_X)$ fixes both coherence phases, whereas
$\Omega_{\rm mod}$ of the main text fixes only $(a,b,c,d,|u|,|v|)$.  For
$\tau\in\Omega_{\rm mod}$ with $X$ part $\tau_X=\Xtwirl(\tau)$, a
Bell-preserving local unitary sets $v=|v|$ while leaving $u$ unchanged,
and it does not change $I_K$, so it carries $\tau_X$ to $\sigma_X(\theta)$
with $\theta=\arg u\in(-\pi,\pi]$.  The mixtures at $\theta$ and $-\theta$ are
local-unitary images of each other at every weight
(Sec.~S\ref{sec:sm-phase-proof}), so $I_K$ is the same at both, and the
representative phase can be taken in $[0,\pi]$.
Theorems~\ref{thm:shadow} and~\ref{thm:phase} then give
$$
I_K(\tau)\subseteq I_K(\tau_X)
\subseteq I_K(\sigma_X(\pi)).
$$
Because $\sigma_X(\pi)\in\Omega_{\rm mod}$, it attains the resulting threshold
upper bound, so
$\max_{\tau\in\Omega_{\rm mod}}\lambda_K(\tau)=\lambda_K(\sigma_X(\pi))$, which
is the corollary of the main text.

\refstepcounter{smsection}
\subsection*{S\arabic{smsection}. Tightness of the \texorpdfstring{$X$}{X}-part threshold bound}
\label{sec:sm-tightness}

The $X$-twirl discards exactly the coherences that connect the even- and
odd-parity sectors.
This section bounds the cost of that missing information.  We use the
main-text threshold gap
$\delta_K(\sigma)=\lambda_K(\sigma_X)-\lambda_K(\sigma)$ and the main-text
interval loss
$\mu_K(\sigma)=\operatorname{meas}[I_K(\sigma_X)\setminus I_K(\sigma)]$.
Both are nonnegative by Theorem~\ref{thm:shadow}.  We also use the distance
$\varepsilon(\sigma):=\tfrac12\|\sigma-\sigma_X\|_1$, which the main text
writes $\varepsilon$.

\subsubsection*{Compatibility class in parity blocks}

Order the computational
basis by parity, first $|00\rangle,|11\rangle$ and then
$|01\rangle,|10\rangle$.  In that order,
\begin{equation}
\sigma=\begin{pmatrix}\sigma_{\mathrm e}&\kappa\\[2pt]
\kappa^\dagger&\sigma_{\mathrm o}\end{pmatrix},
\qquad
\Xtwirl(\sigma)=\begin{pmatrix}\sigma_{\mathrm e}&0\\[2pt]
0&\sigma_{\mathrm o}\end{pmatrix},
\label{eq:parity-blocks}
\end{equation}
so the $X$-twirl keeps the even- and odd-parity blocks
$\sigma_{\mathrm e}=\left(\begin{smallmatrix}a&u\\ \bar u&d\end{smallmatrix}\right)$
and
$\sigma_{\mathrm o}=\left(\begin{smallmatrix}b&v\\ \bar v&c\end{smallmatrix}\right)$
of Eq.~\eqref{eq:xstate}.  It deletes the block $\kappa$ that connects the two
sectors.  The seven expectation values specify $\sigma_{\mathrm e}$ and
$\sigma_{\mathrm o}$ completely.  The remaining block $\kappa$ is constrained
by positivity of the full state.

\begin{lemma}[parity-block positivity]
Let $\sigma_{\mathrm e},\sigma_{\mathrm o}\succeq0$.  The block matrix
$\sigma$ in Eq.~\eqref{eq:parity-blocks} is positive semidefinite if and only
if its interparity block has the form
\begin{equation}
\kappa=\sigma_{\mathrm e}^{1/2}R\,\sigma_{\mathrm o}^{1/2},
\qquad R\in\mathbb C^{2\times2},\qquad \|R\|_\infty\le1 .
\label{eq:contraction-param}
\end{equation}
\end{lemma}

\textit{Proof.}
For sufficiency, take any $R$ with $\|R\|_\infty\leq1$ and $\kappa$ as in
Eq.~\eqref{eq:contraction-param}.  Then
$$
\sigma=
\begin{pmatrix}
\sigma_{\mathrm e}^{1/2}&0\\
0&\sigma_{\mathrm o}^{1/2}
\end{pmatrix}
\begin{pmatrix}
I&R\\
R^\dagger&I
\end{pmatrix}
\begin{pmatrix}
\sigma_{\mathrm e}^{1/2}&0\\
0&\sigma_{\mathrm o}^{1/2}
\end{pmatrix}.
$$
The middle matrix is positive semidefinite when
$\|R\|_\infty\leq1$, since its Schur complement is
$I-R^\dagger R\succeq0$.  Hence, any such $R$ gives a positive semidefinite
$\sigma$.

For necessity, suppose that $\sigma$ is positive semidefinite.  If both
parity blocks are invertible, set
$R=\sigma_{\mathrm e}^{-1/2}\kappa\,\sigma_{\mathrm o}^{-1/2}$.  The Schur
complement of $\sigma_{\mathrm e}$ is
$$
\sigma_{\mathrm o}-\kappa^\dagger\sigma_{\mathrm e}^{-1}\kappa
=\sigma_{\mathrm o}^{1/2}(I-R^\dagger R)\sigma_{\mathrm o}^{1/2}\succeq0.
$$
Because $\sigma_{\mathrm o}$ is positive definite, congruence by
$\sigma_{\mathrm o}^{-1/2}$ makes this inequality equivalent to
$I-R^\dagger R\succeq0$, and hence to $\|R\|_\infty\leq1$.  Thus,
Eq.~\eqref{eq:contraction-param} holds.

It remains to prove necessity when a parity block is singular.  For
$t>0$, define
$$
\sigma_t:=\sigma+tI_4
=\begin{pmatrix}
\sigma_{\mathrm e}+tI&\kappa\\
\kappa^\dagger&\sigma_{\mathrm o}+tI
\end{pmatrix}\succeq0.
$$
Both parity blocks of $\sigma_t$ are invertible.  Applying the invertible
case to $\sigma_t$ gives
$$
R_t:=(\sigma_{\mathrm e}+tI)^{-1/2}\kappa
(\sigma_{\mathrm o}+tI)^{-1/2},
\qquad \|R_t\|_\infty\leq1.
$$
The closed unit ball in $\mathbb C^{2\times2}$ is compact.  Therefore,
there is a sequence $t_n\downarrow0$ for which $R_{t_n}\to R$ and
$\|R\|_\infty\leq1$.  For every $n$,
$$
\kappa
=(\sigma_{\mathrm e}+t_nI)^{1/2}R_{t_n}
(\sigma_{\mathrm o}+t_nI)^{1/2},
$$
and taking $n\to\infty$ gives
$\kappa=\sigma_{\mathrm e}^{1/2}R\,\sigma_{\mathrm o}^{1/2}$.
This proves necessity for singular parity blocks and completes the
equivalence in Eq.~\eqref{eq:contraction-param}.~$\square$

\subsubsection*{Trace-distance radius of the compatibility class}

\textit{The distance for one noise state.}  The difference
$\sigma-\sigma_X$ contains only the off-diagonal blocks in
Eq.~\eqref{eq:parity-blocks}.  Its eigenvalues are
$\pm s_1(\kappa)$ and $\pm s_2(\kappa)$, where $s_1$ and $s_2$ are the
singular values of $\kappa$.  The sum $s_1(\kappa)+s_2(\kappa)$ is the nuclear (trace) norm
$\|\kappa\|_*:=\Tr\sqrt{\kappa^\dagger\kappa}$.  Therefore,
\begin{equation}
\varepsilon(\sigma)
=s_1(\kappa)+s_2(\kappa)=\|\kappa\|_* .
\label{eq:nuclear-distance}
\end{equation}
Both Bell mixtures contain the same Bell-state contribution, so
\begin{equation}
\tfrac12\bigl\|\rho_\lambda(\sigma)
-\rho_\lambda(\sigma_X)\bigr\|_1
=(1-\lambda)\varepsilon(\sigma).
\label{eq:line-distance}
\end{equation}
The uncertainty caused by the unknown coherences is largest at the noise
endpoint and falls linearly to zero at $\PhiP$.  If $\sigma$ is itself of $X$
form, then $\kappa=0$ and $\sigma=\sigma_X$, the two mixing lines coincide,
and $\delta_K(\sigma)=\mu_K(\sigma)=0$ for every $K$.

\textit{The exact worst-compatible distance.}  The value
$\varepsilon(\sigma)$ is not determined by the seven expectation values
because they do not fix $\kappa$.  They do, however, determine the exact
maximum of $\varepsilon$ over the complete compatibility class, and
Eqs.~\eqref{eq:contraction-param} and~\eqref{eq:nuclear-distance} give
$$
\begin{aligned}
\varepsilon_{\max}(\sigma_X)
&:=\max_{\tau\in\Omega(\sigma_X)}
\tfrac12\|\tau-\sigma_X\|_1\\
&=\max_{\substack{
\kappa=\sigma_{\mathrm e}^{1/2}R\sigma_{\mathrm o}^{1/2}\\
\|R\|_\infty\leq1}}\|\kappa\|_*.
\end{aligned}
$$
To see why a unitary $R$ suffices, we write the singular-value
decomposition of any allowed $R$ as
$$
\begin{gathered}
R=V\operatorname{diag}(r_1,r_2)W^\dagger,
\qquad 0\leq r_i\leq1,
\qquad \varphi_i:=\arccos r_i,\\
U_\pm:=V\operatorname{diag}
\bigl(e^{\pm i\varphi_1},e^{\pm i\varphi_2}\bigr)W^\dagger,
\qquad R=\tfrac12(U_++U_-).
\end{gathered}
$$
The matrices $U_\pm$ are unitary.  Write
$f(R):=\|\sigma_{\mathrm e}^{1/2}R\sigma_{\mathrm o}^{1/2}\|_*$.  Convexity
of the nuclear norm gives
$$
f(R)\leq\tfrac12\bigl[f(U_+)+f(U_-)\bigr]
\leq\max\{f(U_+),f(U_-)\}.
$$
Taking the maximum over all allowed $R$ gives
$$
\max_{\|R\|_\infty\leq1}f(R)
\leq\max_{U^\dagger U=I}f(U).
$$
Every unitary satisfies $\|U\|_\infty=1$, so the reverse inequality also
holds.  Therefore,
$$
\max_{\|R\|_\infty\leq1}f(R)
=\max_{U^\dagger U=I}f(U).
$$
Let $\alpha_+\geq\alpha_-$ and $\beta_+\geq\beta_-$ be the eigenvalues
of $\sigma_{\mathrm e}$ and $\sigma_{\mathrm o}$, respectively:
$$
\begin{aligned}
\alpha_\pm&=\tfrac12\bigl[a+d
\pm\sqrt{(a-d)^2+4|u|^2}\bigr],\\
\beta_\pm&=\tfrac12\bigl[b+c
\pm\sqrt{(b-c)^2+4|v|^2}\bigr].
\end{aligned}
$$
For arbitrary $n\times n$ matrices $G$ and $H$, von Neumann's trace
inequality states that
$$
|\Tr(G^\dagger H)|\leq\sum_{i=1}^n s_i(G)s_i(H),
$$
where $s_i(G)$ and $s_i(H)$ denote the singular values in decreasing
order.  We apply the inequality with $G$$=\sigma_{\mathrm o}$ and
$H$$=R^\dagger\sigma_{\mathrm e}R$.  Both matrices are positive semidefinite,
and unitary conjugation preserves the eigenvalues of $\sigma_{\mathrm e}$.
Their singular values are $(\beta_+,\beta_-)$ and $(\alpha_+,\alpha_-)$,
respectively.  Hence,
$$
\Tr(\sigma_{\mathrm o}R^\dagger\sigma_{\mathrm e}R)
\leq\alpha_+\beta_++\alpha_-\beta_-.
$$
Choosing $R$ to map the eigenvectors of $\sigma_{\mathrm o}$ to those of
$\sigma_{\mathrm e}$ in decreasing eigenvalue order attains equality.  Define
$M_R:=\sigma_{\mathrm e}^{1/2}R\sigma_{\mathrm o}^{1/2}$.  For a
$2\times2$ matrix, the square of the nuclear norm is the sum of the squared
singular values plus twice their product.  Therefore,
$$
\begin{aligned}
\|M_R\|_*^2
&=\Tr(M_R^\dagger M_R)+2|\det M_R|\\
&=\Tr(\sigma_{\mathrm o}R^\dagger\sigma_{\mathrm e}R)
+2\sqrt{\alpha_+\alpha_-\beta_+\beta_-}\\
&\leq\alpha_+\beta_++\alpha_-\beta_-
+2\sqrt{\alpha_+\alpha_-\beta_+\beta_-}\\
&=\bigl(\sqrt{\alpha_+\beta_+}
+\sqrt{\alpha_-\beta_-}\bigr)^2.
\end{aligned}
$$
The second equality uses $|\det R|=1$.  Choosing $R$ as above saturates
the upper bound.  Taking the square root, we obtain
\begin{equation}
\varepsilon_{\max}(\sigma_X)
=\sqrt{\alpha_+\beta_+}+\sqrt{\alpha_-\beta_-}.
\label{eq:fiber-radius}
\end{equation}
By the Cauchy--Schwarz inequality, this radius is at most
$\sqrt{(a+d)(b+c)}\le\tfrac12$.
Because the choice of $R$ above attains the bound,
Eq.~\eqref{eq:fiber-radius} gives the exact maximum compatible with the seven
expectation values.  Equation~\eqref{eq:fiber-radius} bounds the distance for every
compatible noise state but does not determine the value attained by an
individual state.

\subsubsection*{Interval and edge bounds}

\textit{General bound on
$I_K(\sigma_X)\setminus I_K(\sigma)$.}
If $\lambda\in I_K(\sigma_X)\setminus I_K(\sigma)$, then
$\rho_\lambda(\sigma_X)\in K$, whereas
$\rho_\lambda(\sigma)\notin K$, so their trace distance is at least the
shortest distance from $\rho_\lambda(\sigma_X)$ to an operator outside $K$.
Let $\mathcal A$
be the set of unit-trace Hermitian operators, regard $K$ as a convex
subset of $\mathcal A$, and define, for $\lambda\in I_K(\sigma_X)$,
\begin{equation}
m(\lambda):=
\inf_{H\in\mathcal A\setminus K}
\tfrac12\bigl\|\rho_\lambda(\sigma_X)-H\bigr\|_1 .
\label{eq:margin-definition}
\end{equation}
Every unit-trace Hermitian operator closer than $m(\lambda)$ to
$\rho_\lambda(\sigma_X)$ therefore lies in $K$.  The main-text distance $\mathcal D_K(\sigma_X)$ equals
$m(0)$, since both infima are over $\mathcal A\setminus K$.  We take the infimum
over this ambient set, rather than over quantum states alone, because that choice makes
$m$ concave along the mixing line, and the edge bounds below need that concavity.  The
larger set can only lower $m$, so it weakens every bound below and never makes
one false.

If $\rho_\lambda(\sigma)\notin K$ while
$\rho_\lambda(\sigma_X)\in K$, then
$\rho_\lambda(\sigma)\in\mathcal A\setminus K$, and therefore
\begin{equation}
m(\lambda)
\le\tfrac12\bigl\|\rho_\lambda(\sigma_X)-\rho_\lambda(\sigma)\bigr\|_1
=(1-\lambda)\tfrac12\|\sigma_X-\sigma\|_1
=(1-\lambda)\varepsilon(\sigma).
\label{eq:margin-displacement}
\end{equation}
Consequently,
\begin{equation}
I_K(\sigma_X)\setminus I_K(\sigma)
\subseteq
\bigl\{\lambda\in I_K(\sigma_X):
m(\lambda)\le(1-\lambda)\varepsilon(\sigma)\bigr\}.
\label{eq:disagreement-set}
\end{equation}
The condition in Eq.~\eqref{eq:disagreement-set} is necessary but not
sufficient, so $m(\lambda)\leq(1-\lambda)\varepsilon(\sigma)$ does not by
itself put $\rho_\lambda(\sigma)$ outside $K$.  Replacing
$\varepsilon(\sigma)$ by $\varepsilon_{\max}(\sigma_X)$ makes the inclusion in
Eq.~\eqref{eq:disagreement-set} uniform over
every $\sigma\in\Omega(\sigma_X)$.  Bounding the interval endpoints needs
the concavity of $m$.

The function $m$ is concave on $I_K(\sigma_X)$.  To see this, take
$\lambda_1,\lambda_2\in I_K(\sigma_X)$ and $t\in[0,1]$.  Choose $r_i\geq0$
with $r_i<m(\lambda_i)$ whenever $m(\lambda_i)>0$, and take $r_i=0$ when
$m(\lambda_i)=0$.  Write $c_i=\rho_{\lambda_i}(\sigma_X)$,
$c=tc_1+(1-t)c_2$, and $\bar r=tr_1+(1-t)r_2$, so that
$c=\rho_{t\lambda_1+(1-t)\lambda_2}(\sigma_X)$.  The trace-norm balls
$$
B_i:=\left\{H\in\mathcal A:
\tfrac12\bigl\|H-c_i\bigr\|_1\leq r_i\right\}
$$
lie in $K$.  Because $K$ is convex, $tB_1+(1-t)B_2\subseteq K$.
Moreover,
$$
tB_1+(1-t)B_2=\left\{H\in\mathcal A:
\tfrac12\|H-c\|_1\leq \bar r\right\}.
$$
For $H_i\in B_i$, the triangle inequality applied to
$tH_1+(1-t)H_2-c=t(H_1-c_1)+(1-t)(H_2-c_2)$ gives
$\tfrac12\|tH_1+(1-t)H_2-c\|_1\leq tr_1+(1-t)r_2=\bar r$.  Conversely, if $\tfrac12\|H-c\|_1\leq \bar r$ with $\bar r>0$, then
$H_i=c_i+(r_i/\bar r)(H-c)$ satisfies $\tfrac12\|H_i-c_i\|_1\leq r_i$ and
$tH_1+(1-t)H_2=H$, while for $\bar r=0$ both sides reduce to $\{c\}$.  Hence, the
ball of radius $\bar r$ about $c$ lies in $K$, so
$m(t\lambda_1+(1-t)\lambda_2)\geq \bar r$.  This holds for every admissible pair
$r_1,r_2$, the left side does not depend on them, and $\sup r_i=m(\lambda_i)$,
so
$$
m\bigl(t\lambda_1+(1-t)\lambda_2\bigr)
\geq t\,m(\lambda_1)+(1-t)m(\lambda_2).
$$

Suppose
\begin{equation}
I_K(\sigma_X)=[\lambda_-,\lambda_+],
\qquad \lambda_-<\lambda_+,
\label{eq:x-interval-general}
\end{equation}
where $\lambda_-=0$ describes an anchored interval and $\lambda_->0$ a detached
interval.  The detached case is the case $\sigma_X\notin K$ of the main
text.  Choose $\lambda_*\in(\lambda_-,\lambda_+)$ and choose either
\begin{equation}
\eta=\varepsilon(\sigma)
\quad\text{or}\quad
\eta=\varepsilon_{\max}(\sigma_X).
\label{eq:eta-choice}
\end{equation}
Write
$$
s(\lambda):=m(\lambda)-(1-\lambda)\eta .
$$
This function is concave, since $m$ is concave and $(1-\lambda)\eta$ is
linear in $\lambda$.
Equation~\eqref{eq:margin-displacement} gives
$\rho_\lambda(\sigma)\in K$ whenever $s(\lambda)>0$.  The condition
\begin{equation}
s_*:=s(\lambda_*)>0
\label{eq:certified-anchor}
\end{equation}
therefore gives $\lambda_*\in I_K(\sigma)$ when $\eta=\varepsilon(\sigma)$,
and $\lambda_*\in I_K(\tau)$ for every $\tau\in\Omega(\sigma_X)$ when
$\eta=\varepsilon_{\max}(\sigma_X)$.

\begin{figure}[t]
\centering
\includegraphics{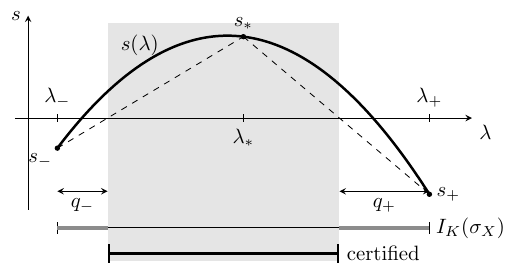}
\caption{The margin $s(\lambda)=m(\lambda)-(1-\lambda)\eta$ on
$I_K(\sigma_X)=[\lambda_-,\lambda_+]$, drawn negative at both ends.  Concavity
keeps the solid curve above each dashed chord.  The chord from
$(\lambda_-,s_-)$ to $(\lambda_*,s_*)$ reaches zero a distance $q_-$ after
$\lambda_-$, and the chord from $(\lambda_*,s_*)$ to $(\lambda_+,s_+)$ reaches
zero a distance $q_+$ before $\lambda_+$.  On the shaded band, both chords are
nonnegative, so $s\geq0$ there and
$[\lambda_-+q_-,\lambda_+-q_+]\subseteq I_K(\sigma)$.  The bars below mark
$I_K(\sigma_X)$ and that inner interval, labeled ``certified''.  Because $I_K(\sigma)$ lies
between those two intervals and is itself an interval, the construction confines its two
endpoints to the two gray segments at the ends of the $I_K(\sigma_X)$ bar, of
lengths $q_-$ and $q_+$.  The zeros of
the curve lie outside the shaded band, so $s\geq0$ on a wider range than the chords
certify.  Consequently, $\lambda_K(\sigma)$ lies between the larger zero of
$s$ and $\lambda_+$, and the other endpoint of $I_K(\sigma)$ lies between
$\lambda_-$ and the smaller zero.  Each of these ranges is shorter than
the corresponding gray segment.}
\label{fig:chords}
\end{figure}

Write $s_-=s(\lambda_-)$ and $s_+=s(\lambda_+)$.  Concavity means that the
graph of $s$ never falls below the straight segment joining two of its points,
called a chord.  The graph therefore lies above the chord from
$(\lambda_-,s_-)$ to $(\lambda_*,s_*)$ and above the chord from
$(\lambda_*,s_*)$ to $(\lambda_+,s_+)$.  The first chord rises from $s_-$ at
$\lambda_-$ to $s_*>0$ at $\lambda_*$.  If $s_-<0$, it crosses zero after a fraction
$-s_-/(s_*-s_-)$ of the distance from $\lambda_-$ to $\lambda_*$, and if
$s_-\geq0$, it stays nonnegative from $\lambda_-$ to $\lambda_*$.  The second chord is read the same
way but in the opposite direction, from $\lambda_+$ back toward $\lambda_*$,
with $s_+$ in place of $s_-$ (Fig.~\ref{fig:chords}).  The two chords therefore reach zero at
$\lambda_-+q_-$ and at $\lambda_+-q_+$, where
\begin{equation}
q_-=
\begin{cases}
0, & s_-\ge0,\\[2pt]
(\lambda_*-\lambda_-)\dfrac{-s_-}{s_*-s_-}, & s_-<0,
\end{cases}
\quad
q_+=
\begin{cases}
0, & s_+\ge0,\\[2pt]
(\lambda_+-\lambda_*)\dfrac{-s_+}{s_*-s_+}, & s_+<0.
\end{cases}
\label{eq:two-edge-losses}
\end{equation}
Since $s$ lies above both chords, $s>0$ strictly between those points, so
the open interval $(\lambda_-+q_-,\lambda_+-q_+)$ lies in $I_K(\sigma)$.  For
closed $K$, continuity of $\lambda\mapsto\rho_\lambda(\sigma)$ makes
$I_K(\sigma)$ closed, and therefore
\begin{equation}
[\lambda_-+q_-,\,\lambda_+-q_+]\subseteq I_K(\sigma).
\label{eq:certified-inner-interval}
\end{equation}

In particular, the threshold gap
$\delta_K(\sigma)=\lambda_K(\sigma_X)-\lambda_K(\sigma)$ and the interval loss
$\mu_K(\sigma)=\operatorname{meas}[I_K(\sigma_X)\setminus I_K(\sigma)]$ obey
\begin{equation}
0\le \delta_K(\sigma)\le q_+,
\qquad
\mu_K(\sigma)\le q_-+q_+.
\label{eq:general-edge-bounds}
\end{equation}
Both bounds are conservative, for two independent reasons visible in
Fig.~\ref{fig:chords}.  Wherever the
concavity is strict, $s$ reaches zero outside $\lambda_-+q_-$ and
$\lambda_+-q_+$.  Beyond that, $s(\lambda)\le0$ does not
put $\rho_\lambda(\sigma)$ outside $K$, so $I_K(\sigma)$ can extend past the
set where $s\ge0$.  Equality in Eq.~\eqref{eq:general-edge-bounds} would
require $s$ to be affine along the chord and $I_K(\sigma)$ to coincide with
that set.
No $\lambda_*$ satisfies Eq.~\eqref{eq:certified-anchor} exactly when
$\eta\ge\sup_\lambda m(\lambda)/(1-\lambda)$ over $I_K(\sigma_X)$.  The
right-hand side of Eq.~\eqref{eq:disagreement-set} is then all of
$I_K(\sigma_X)$, so the inclusion holds automatically and says nothing about
$I_K(\sigma_X)\setminus I_K(\sigma)$, which may still be large.  No edge bound
follows.  Only
a smaller $\eta$ helps, either $\varepsilon(\sigma)$ in place of
$\varepsilon_{\max}(\sigma_X)$ when the noise state itself is known, or
$\varepsilon(\sigma)=0$ when the noise is known to preserve the $X$ form.

\begin{figure}[t]
\centering
\includegraphics{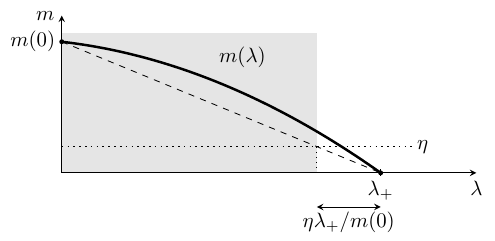}
\caption{The anchored case $\sigma_X\in K$, with $I_K(\sigma_X)=[0,\lambda_+]$.
The hypothesis $m(0)>0$ and the derived $m(\lambda_+)=0$ fix the dashed chord,
and concavity keeps $m$ above it, which is
Eq.~\eqref{eq:margin-slope}.  The dotted line is the constant $\eta$, which
replaces $(1-\lambda)\eta$.  It meets the chord at
$\lambda_+-\eta\lambda_+/m(0)$, so the shaded interval lies in $I_K(\sigma)$
and Eq.~\eqref{eq:margin-gap} follows.  Both the chord and the constant $\eta$ are
conservative, since the chord lies below $m$ and $\eta$ lies above
$(1-\lambda)\eta$, so the true crossing lies to the right of the shaded edge.}
\label{fig:anchored}
\end{figure}

The bound quoted in the main text is the anchored special case, in which
$\sigma_X\in K$.
Let $I_K(\sigma_X)=[0,\lambda_+]$ with $0<\lambda_+<1$ and $m(0)>0$.  Then
$m(\lambda_+)=0$, because $\rho_\lambda(\sigma_X)$ lies outside $K$ for every
$\lambda>\lambda_+$, and those states approach $\rho_{\lambda_+}(\sigma_X)$ as
$\lambda$ decreases to $\lambda_+$.  The chord from $(0,m(0))$ to
$(\lambda_+,0)$ therefore gives (Fig.~\ref{fig:anchored})
\begin{equation}
m(\lambda)\ge\frac{m(0)}{\lambda_+}(\lambda_+-\lambda).
\label{eq:margin-slope}
\end{equation}
At a weight $\lambda\in I_K(\sigma_X)\setminus I_K(\sigma)$,
Eq.~\eqref{eq:disagreement-set} bounds $m(\lambda)$ from above and
Eq.~\eqref{eq:margin-slope} bounds it from below, so
$$
\frac{m(0)}{\lambda_+}(\lambda_+-\lambda)\le m(\lambda)
\le(1-\lambda)\eta\le\eta ,
$$
and therefore $\lambda_+-\lambda\le\eta\,\lambda_+/m(0)$.  Equivalently,
\begin{equation}
I_K(\sigma_X)\setminus I_K(\sigma)\subseteq
\Bigl[\lambda_+-\frac{\eta\,\lambda_+}{m(0)},\;\lambda_+\Bigr] .
\label{eq:top-stretch}
\end{equation}
The two intervals therefore coincide below
$\lambda_+-\eta\,\lambda_+/m(0)$, so that value bounds $\lambda_K(\sigma)$ from
below, and the right-hand side has length $\eta\,\lambda_+/m(0)$, which bounds
$\mu_K(\sigma)$.  Hence,
\begin{equation}
\max\{\delta_K(\sigma),\mu_K(\sigma)\}
\le\frac{\eta\,\lambda_+}{m(0)}.
\label{eq:margin-gap}
\end{equation}
If $\eta\ge m(0)$, Eq.~\eqref{eq:margin-gap} is no stronger than the
trivial bound by the length $\lambda_+$.

The main text quotes Eq.~\eqref{eq:margin-gap} under the single hypothesis
$\mathcal D_K(\sigma_X)=m(0)>0$, and that hypothesis supplies the rest.  A
positive $m(0)$ places $\sigma_X$ in the interior of $K$, so the interval
is anchored and $\lambda_+>0$.  The remaining requirement $\lambda_+<1$ holds whenever $K$ is closed, since
the supremum is then attained and $\lambda_+=1$ would put $\PhiP$ itself in
$K$, whereas $\PhiP$ has every ability considered here.  Every set listed in Sec.~S\ref{sec:sm-phase-proof} is a closed set of states, as shown there.

Equation~\eqref{eq:margin-gap} assumes nothing about $\sigma$ beyond its $X$
part, and Eq.~\eqref{eq:top-stretch} shows what that costs.  The right-hand
side of Eq.~\eqref{eq:top-stretch} avoids $\lambda=0$ exactly when $\eta<m(0)$, so in that range $\sigma\in K$
follows rather than being assumed, $I_K(\sigma)$ is anchored as well, and
$\mu_K(\sigma)=\delta_K(\sigma)$.  For $\eta=\varepsilon_{\max}(\sigma_X)$, the
same holds for every compatible noise state.  When $\eta\ge m(0)$, that side
covers all of $[0,\lambda_+]$ and nothing is left.  A noise state that already
has the ability therefore lies outside the reach of
Eq.~\eqref{eq:margin-gap}.  Its own interval $I_K(\sigma)$ is then empty or
starts above zero, while $I_K(\sigma_X)$ can stay anchored, because Theorem~\ref{thm:shadow}
carries $\sigma_X\notin K$ to $\sigma\notin K$ and not the converse.  The bound
that applies there is Eq.~\eqref{eq:general-edge-bounds}, which assumes nothing
about $\lambda_-$ and instead needs a weight $\lambda_*$ at which $s$ is
positive, as in Eq.~\eqref{eq:certified-anchor}.

\refstepcounter{smsection}
\subsection*{S\arabic{smsection}. Coherence-reduction channels}
\label{sec:sm-contraction}

\subsubsection*{Coherence-reduction lemma}

\textit{Lemma.}  Let $\rho$ and $\rho'$ be two-qubit $X$ states with
the same populations $(a,b,c,d)$, the same odd-sector coherence
modulus $|w|$, and even-sector coherence moduli $q\ge q'$.  Then
$\rho'$ is a convex mixture of local-unitary images of $\rho$.

\textit{Proof.}  Write the even-sector coherences of $\rho$ and $\rho'$ as
$z=qe^{i\chi}$ and $z'=q'e^{i\chi'}$, and their odd-sector coherences as
$w=|w|e^{i\psi}$ and $w'=|w|e^{i\psi'}$.  If a modulus vanishes, its phase
is arbitrary.  If $q=0$, then $q'=0$, and $W(\psi-\psi',0)$ carries $\rho$ to
$\rho'$.  Otherwise, pick $\varphi\in[0,\pi/4]$ with $\cos2\varphi=q'/q$ and set
$$
\alpha=\tfrac12\bigl[(\chi-\chi')+(\psi-\psi')\bigr],
\quad
\beta=\tfrac12\bigl[(\chi-\chi')-(\psi-\psi')\bigr],
\quad
W_\pm=W(\alpha\pm\varphi,\,\beta\pm\varphi).
$$
Each $W_\pm$ is diagonal, so it fixes the populations while multiplying $z$ by
$e^{-i(\alpha+\beta\pm2\varphi)}$ and $w$ by $e^{-i(\alpha-\beta)}$.  Here,
$\alpha+\beta=\chi-\chi'$ and $\alpha-\beta=\psi-\psi'$, and the two
even-sector factors average to $e^{-i(\alpha+\beta)}\cos2\varphi$, so the
equal-weight mixture
$\tfrac12 W_+\rho W_+^\dagger+\tfrac12 W_-\rho W_-^\dagger$ has even-sector
coherence $q\cos(2\varphi)e^{i\chi'}=z'$, odd-sector coherence
$|w|e^{i\psi'}=w'$, and the populations of $\rho$.  The mixture therefore
equals $\rho'$, so $\rho'$ is an equal-weight mixture of two local-unitary images of
$\rho$.~$\square$

\subsubsection*{Applying the lemma to the mixing family}

Fix $\lambda$ and $0\le\theta\le\theta'\le\pi$.  The mixtures
$\rho_\lambda(\sigma_X(\theta))$ and $\rho_\lambda(\sigma_X(\theta'))$ have the
same populations and the same odd-sector coherence $(1-\lambda)v$, and their
even-sector coherences have moduli $q_\lambda(\theta')\le q_\lambda(\theta)$.
The lemma therefore writes $\rho_\lambda(\sigma_X(\theta'))$ as an equal-weight
mixture of two local-unitary images of $\rho_\lambda(\sigma_X(\theta))$, with
$\cos2\varphi=q_\lambda(\theta')/q_\lambda(\theta)$ and, because the
odd-sector coherences are the same, with
$\alpha=\beta=\tfrac12(\chi-\chi')$, where $\chi$ and $\chi'$ are the phases of
the two even-sector coherences.  The
parties realize the mixture by tossing a fair coin and applying the phase gates
its outcome selects.

The conversion runs one way.  A mixture $\sum_i p_i W(\alpha_i,\beta_i)\,
\rho\,W(\alpha_i,\beta_i)^\dagger$ of local phase unitaries multiplies the
even-sector coherence by $\sum_i p_i e^{-i(\alpha_i+\beta_i)}$, an average of
numbers of modulus one, so the factor has modulus at most one and $q_\lambda$
can fall but never rise.  By the expression for $q_\lambda(\theta)^2$ in the
main text, $q_\lambda(\theta)>q_\lambda(\theta')$ whenever $0<\lambda<1$,
$|u|>0$, and $\theta<\theta'$, so the reverse conversion would need a factor of
modulus above one, which no such mixture supplies.

\subsubsection*{Fixed-population coherence ordering}

This part proves the extension of Theorem~\ref{thm:phase} to the coherence
moduli, stated in the main text.  The hypotheses of that extension allow
the odd-sector coherence
modulus to drop as well, which the lemma does not cover.

\textit{Proposition.}  Let $\sigma_X$ and $\sigma'_X$ be $X$ states with
the same populations, with even-sector coherences $u=re^{i\theta}$ and
$u'=r'e^{i\theta'}$ and odd-sector coherences $v$ and $v'$, where
$\theta,\theta'\in[0,\pi]$.  If $r'\le r$, $r'\cos\theta'\le r\cos\theta$, and
$|v'|\le|v|$, then for every convex ability-absence set $K$ closed under local
unitaries,
$$
I_K(\sigma_X)\subseteq I_K(\sigma'_X),
\qquad
\lambda_K(\sigma_X)\le\lambda_K(\sigma'_X).
$$

\textit{Proof.}  We first extend the conversion of the lemma to shrink
both coherences at once.  Let $\rho$ and $\rho'$ be $X$ states with
the same populations, with even-sector coherences $z$ and $z'$ and
odd-sector coherences $w$ and $w'$.  Suppose
$z'=\eta_u z$ and $w'=\eta_v w$, with $|\eta_u|\le1$ and $|\eta_v|\le1$.  Then
$\rho'$ is an equal-weight mixture of four local-unitary images of $\rho$.  To see
this, write $\eta_u=|\eta_u|e^{i\gamma_u}$ and $\eta_v=|\eta_v|e^{i\gamma_v}$,
and set $x_\pm=-\gamma_u\pm\arccos|\eta_u|$ and
$y_\pm=-\gamma_v\pm\arccos|\eta_v|$, so that
$\tfrac12(e^{-ix_+}+e^{-ix_-})=e^{i\gamma_u}|\eta_u|=\eta_u$ and
$\tfrac12(e^{-iy_+}+e^{-iy_-})=\eta_v$.  Since $W(\alpha,\beta)$ multiplies $z$
by $e^{-i(\alpha+\beta)}$ and $w$ by $e^{-i(\alpha-\beta)}$, the choice
$$
W_{st}=W\Bigl(\frac{x_s+y_t}{2},\frac{x_s-y_t}{2}\Bigr),
\qquad s,t\in\{+,-\} ,
$$
multiplies $z$ by $e^{-ix_s}$ and $w$ by $e^{-iy_t}$.  Averaging the four
images $W_{st}\,\rho\,W_{st}^\dagger$ with weight $\tfrac14$ therefore
multiplies $z$ by $\eta_u$ and $w$ by $\eta_v$ and leaves the populations
unchanged, which gives $\rho'$.

We now apply this conversion at each Bell weight.  The mixtures
$\rho_\lambda(\sigma_X)$ and $\rho_\lambda(\sigma'_X)$ have the same
populations, even-sector coherences $z_\lambda(u)$ and $z_\lambda(u')$ with
$z_\lambda(u)=\tfrac{\lambda}{2}+(1-\lambda)u$, and odd-sector coherences
$(1-\lambda)v$ and $(1-\lambda)v'$.  Direct computation gives
\begin{equation}
|z_\lambda(u')|^{2}-|z_\lambda(u)|^{2}
=(1-\lambda)\bigl[\lambda(r'\cos\theta'-r\cos\theta)
+(1-\lambda)(r'^2-r^2)\bigr] .
\label{eq:coherence-order}
\end{equation}
The sum in the bracket is affine in $\lambda$, with value $r'^2-r^2$ at $\lambda=0$
and $r'\cos\theta'-r\cos\theta$ at $\lambda=1$, so it is nonpositive on
$[0,1]$ if and only if $r'\le r$ and $r'\cos\theta'\le r\cos\theta$.  Under the first two hypotheses of the Proposition, $|z_\lambda(u')|\le|z_\lambda(u)|$
holds at every Bell weight.  At each fixed Bell weight, this inequality gives
$|\eta_u|\le1$ for the factor $\eta_u=z_\lambda(u')/z_\lambda(u)$, and the
third hypothesis gives $|\eta_v|\le1$ for $\eta_v=v'/v$.  The conversion we
just described with four local unitaries then
writes $\rho_\lambda(\sigma'_X)$ as an equal-weight mixture of four local-unitary
images of $\rho_\lambda(\sigma_X)$.  If $z_\lambda(u)=0$, the factor $\eta_u$ is undefined, but the
inequality above forces $z_\lambda(u')=0$, so the even-sector coherence is
zero in both mixtures and needs no conversion.
The same holds for $\eta_v$ when $v=0$, since $|v'|\le|v|$ forces $v'=0$.  If
$\rho_\lambda(\sigma_X)\in K$, convexity and local-unitary closure of $K$
then give $\rho_\lambda(\sigma'_X)\in K$, as in the proof of
Theorem~\ref{thm:phase}.  This proves the interval inclusion, and taking
suprema gives the threshold ordering.~$\square$

The remarks below delimit the reach of the Proposition.  The phases $x_\pm$ of the proof depend on $\lambda$ through
$\eta_u=z_\lambda(u')/z_\lambda(u)$, so the conversion depends on
$\lambda$ and on both states.
The condition $r'\le r$ alone does not give
$|z_\lambda(u')|\le|z_\lambda(u)|$ at every Bell weight.  The Proposition
therefore pairs it with $r'\cos\theta'\le r\cos\theta$.
Take $u_1=r_1e^{i\theta}$ and $u_2=r_2e^{i\theta}$ with $r_1>r_2\ge0$.
Equation~\eqref{eq:coherence-order} at $\theta'=\theta$, once we pull out the
factor $r_1-r_2$, gives
$$
|z_\lambda(u_1)|^{2}-|z_\lambda(u_2)|^{2}
=(r_1-r_2)(1-\lambda)\bigl[(1-\lambda)(r_1+r_2)+\lambda\cos\theta\bigr] .
$$
For $\cos\theta\ge0$, this is nonnegative at every weight, so
$|z_\lambda(u_2)|\le|z_\lambda(u_1)|$ on $0\le\theta\le\pi/2$.  For $\cos\theta<0$, it changes sign at
$$
\lambda_\times(\theta)=\frac{r_1+r_2}{r_1+r_2-\cos\theta}\in(0,1) .
$$
Below $\lambda_\times(\theta)$ the state with the larger $r$ has the larger
even-sector modulus, and above that weight the state with the smaller $r$
does, so neither state has the larger modulus at every Bell weight.  These are
the incomparable cases mentioned in the main text.
The odd-sector hypothesis $|v'|\le|v|$ needs no such pairing.
Reducing $|v|$ lowers the odd-sector coherence
$(1-\lambda)v$ at every weight, since $\PhiP$ carries no odd-sector coherence.

\refstepcounter{smsection}
\subsection*{S\arabic{smsection}. Phase monotonicity}
\label{sec:sm-phase-proof}

The restriction to $\theta\in[0,\pi]$ in the main text rests on a
local-unitary equivalence.  The states at $\pm\theta$ are local-unitary
images of each other at every weight, by the gauge
$W(\arg z_\lambda,\arg z_\lambda)$ that carries $z_\lambda$ to
$\bar z_\lambda$ and fixes the odd-sector coherence.  That gauge angle
depends on $\arg z_\lambda$, so the equivalence is pointwise in $\lambda$.

\subsubsection*{Convexity and local-unitary closure}

Convexity and local-unitary closure are standard for the abilities named
in the main text, and we record the reasons briefly.  The separable set is
convex and closed under local unitaries.  The teleportation-useless set
$\{f\le\tfrac12\}$ is a sublevel set of the fully entangled fraction
$f(\rho)=\max_{|\beta\rangle}\langle\beta|\rho|\beta\rangle$, which is a maximum
of linear functionals and hence convex, and local unitaries permute the maximally
entangled vectors.  The CHSH-local set $\{M\le1\}$ and the CJWR-satisfying sets $\{F_n\le1\}$ are
convex and closed under local
unitaries~\cite{HorodeckiCHSH1995,CostaAngelo2016,OperationalLayersCompanion},
where $M(\rho)=s_1^{2}+s_2^{2}$ for the two largest singular values of the
correlation tensor $T_\rho$ and $F_n=(\sum_{i=1}^{n}s_i^{2})^{1/2}$ for
$n\in\{2,3\}$.  The unsteerable sets, in each direction and for projective measurements
or POVMs, and the Bell-local sets are convex and closed under local
unitaries~\cite{WisemanJonesDoherty2007,BrunnerEtAl2014}.

Each of these sets is also a closed set of states.  That is a different
property from closure under local unitaries, and it is the one that makes
$I_K(\sigma)$ of Eq.~\eqref{eq:threshold} a closed interval and forces
$\lambda_+<1$ in Eq.~\eqref{eq:margin-gap}.  For the teleportation-useless,
CHSH-local, and CJWR-satisfying sets, it follows from continuity, since $f$,
$M$, and $F_n$ are continuous in $\rho$ and a sublevel set of a continuous
function is closed.  The separable set is the convex hull of the compact set of
pure product states, and in finite dimension such a hull is compact.  The unsteerable sets are closed as well: for a fixed measurement assemblage the local-hidden-state constraints are closed conditions, and the unsteerable set is their intersection over all assemblages [Ref.~\cite{NguyenNguyenGuhne2019}, Proposition~22 proves the corresponding statement for the level sets of the critical radius, relatively in the physical states used here].  So are the Bell-local sets [Ref.~\cite{BrunnerEtAl2014}, Sec.~II.B].

Unlike the other sets here, the unsteerable and Bell-local sets carry no
closed-form membership criterion for two-qubit states, including those in the
$X$ family.  Their thresholds
are therefore beyond the direct computation used in
Ref.~\cite{OperationalLayersCompanion}.

The remaining set consists of the states where the BB84 key-rate bound
specified next is not positive.

\subsubsection*{The measurement-basis-optimized asymptotic lower bound for
one-way BB84 with CSS codes}

We first specify the operational meaning of the bound.  In the
entanglement-based realization of BB84, parties $A$ and $B$ share a two-qubit
pair in state $\rho$ in each round and independently measure in either
the $Z$ or $X$ basis~\cite{BennettBrassardMermin1992}.  After announcing their
basis choices, they keep the rounds in which the bases match.  They take
the key from the rounds measured in $Z$.  The parties announce their outcomes
in a random subset of those rounds.  The fraction of outcome disagreement in
that subset estimates the bit error rate $e_b$, the probability that the two
$Z$-basis outcomes disagree.  The outcomes of all the rounds with matched
$X$ basis are announced, since none of them enters the key.  The fraction of
outcome disagreement in the $X$ basis estimates the phase error rate $e_p$.
Reference~\cite{ShorPreskill2000} calls the announced outcomes the check
bits.  They are used up by the comparison and are excluded from the key.
The $Z$-basis outcomes that were not announced stay secret and form the raw
key, one string held by each party.

Because the check bits were a random sample of the $Z$ rounds, the two raw
key strings disagree approximately in a fraction $e_b$ of their positions.  A
key is useless unless both parties hold the same string, so they must first
use an error correction code to make the strings identical.  This step is
called reconciliation, and Ref.~\cite{ShorPreskill2000} performs it with a
classical linear code.  To apply the code, $A$ sends $B$ a set of parity
bits computed from its raw key string.  These parity bits are public and will
be discarded.  Then, $B$ uses them to correct its own string so that it is
identical to $A$'s remaining raw key string.  For bits that flip
independently with probability $e_b$, Shannon's channel coding theorem
requires the number of parity bits to be at least a fraction $h(e_b)$ of the
raw key string, where $h(e)=-e\log_2e-(1-e)\log_2(1-e)$ is the binary entropy,
and classical linear codes attain that value~\cite{ShorPreskill2000}.  The
rate below charges that fraction.  The
parties fix the code in advance, and they abort and discard the strings if the
estimated $e_b$ exceeds the error rate that the code can correct.

Reconciliation leaves the parties with a common string but says nothing
about what a third party knows of that string.  The phase error rate controls
that knowledge.  If the pairs were exactly $\PhiP$, the $Z$-basis outcomes
would be random and correlated with nothing outside the pair, so an eavesdropper would
know nothing.  To see what a phase error does, write the four Bell states
$\Phi^\pm=(|00\rangle\pm|11\rangle)/\sqrt2$ and $\Psi^\pm=(|01\rangle\pm|10\rangle)/\sqrt2$,
which form an orthonormal basis, and let
$p_{\Phi^\pm}=\langle\Phi^\pm|\rho|\Phi^\pm\rangle$ and
$p_{\Psi^\pm}=\langle\Psi^\pm|\rho|\Psi^\pm\rangle$ be the diagonal elements of
$\rho$ in that basis.  A phase flip $Z\otimes I_2$ turns $\PhiP$ into
$\Phi^-$: it leaves the $Z$-basis outcomes agreeing but makes the $X$-basis
outcomes disagree.  Since $Z\otimes Z$ has eigenvalue $+1$ on $\Phi^\pm$
and $-1$ on $\Psi^\pm$, and $X\otimes X$ has eigenvalue $+1$ on $\Phi^+$
and $\Psi^+$ and $-1$ on $\Phi^-$ and $\Psi^-$, the two rates and the
overlap $F=\langle\PhiP|\rho|\PhiP\rangle$ are $e_b=p_{\Psi^+}+p_{\Psi^-}$, $e_p=p_{\Phi^-}+p_{\Psi^-}$, and
$1-F=p_{\Phi^-}+p_{\Psi^+}+p_{\Psi^-}$.  These give $e_p\le1-F\le e_b+e_p$
for every two-qubit state.  A large $e_p$ therefore forces the pairs far from
$\PhiP$, and the pairs are close only when both rates are small.
Reference~\cite{ShorPreskill2000} first proves security for a protocol in
which the parties correct the bit and phase errors on the qubit pairs
themselves before measuring them.  The correction uses a quantum code of the type introduced by Calderbank, Shor, and
Steane (CSS), and correcting the phase errors costs a further fraction
$h(e_p)$ of the pairs.  After that correction, the shared state is, with high
probability, at a distance from a product of $\PhiP$ pairs that falls
exponentially in the number of pairs.  The information an eavesdropper holds
about the key is bounded in terms of that distance, and is therefore
exponentially small in the number of pairs as well.
Reference~\cite{ShorPreskill2000} then shows that this correction never has to
be carried out.  The parties measure first and afterwards shorten the
reconciled string by $h(e_p)$ bits per raw key bit, a step called privacy
amplification.  Its output is the key of the protocol above.  An eavesdropper's
information about that key is exponentially small in the number of pairs, the
same guarantee proved for the protocol that corrects the errors on the qubit
pairs.

Each outcome is labeled by $\pm1$, and for a two-qubit state $\rho$, the two
rates are read off its correlators,
$$
e_b=\frac{1-\langle Z\otimes Z\rangle_\rho}{2},
\qquad
e_p=\frac{1-\langle X\otimes X\rangle_\rho}{2},
$$
and both vanish on $\PhiP$, whose outcomes are perfectly correlated in
both bases.

Both reconciliation and privacy amplification act only on the measured
outcomes, and every message about the key goes from $A$ to $B$, and nothing is
sent back.  A protocol whose key is processed in that one direction is called
one-way, and the rate below is what it yields.  When the two error
rates are equal, Ref.~\cite{ShorPreskill2000} states that a fraction $1-2h(e)$
of each raw key bit stays secret.  When they differ, the same argument gives,
for the pair of measurement bases fixed above, the lower bound on that
fraction
$$
r=1-h(e_b)-h(e_p),
$$
in which $h(e_b)$ is the cost of reconciliation and $h(e_p)$ is the cost of
privacy amplification.

We next optimize over the local measurement bases.  Let
$a_1,a_2\in\mathbb R^3$ and $b_1,b_2\in\mathbb R^3$ be orthonormal Bloch
directions for $A$ and $B$, respectively, and define the correlation
tensor $T_\rho=(T_{jk})$ by
$T_{jk}=\langle\sigma_j\otimes\sigma_k\rangle_\rho$.  For basis pair $i$,
let $c_i=a_i^{\mathsf T}T_\rho b_i$.  Each party records $\pm1$, and $c_i$ is
the expectation of the product of the two outcomes, so the two outcomes
disagree with probability $(1-c_i)/2$.  Which outcome a party calls $+1$ is a
convention, and reversing it at either party sends $c_i$ to $-c_i$.  The
parties read the disagreement fraction off the check bits.  If it is above one
half, one of them flips the sign of all its outcomes, which brings the
disagreement fraction below one half.
This is the convention with $c_i\ge0$, so the error rate is
$e_i=(1-|c_i|)/2$.
If $s_1\ge s_2\ge s_3$ are the singular values of
$T_\rho$, then
\begin{equation}
r_{\rm opt}
=\max_{\substack{a_1\perp a_2\\ b_1\perp b_2}}
 \left[1-h\!\left(\tfrac{1-|c_1|}{2}\right)
          -h\!\left(\tfrac{1-|c_2|}{2}\right)\right]
=1-h\!\left(\tfrac{1-s_1}{2}\right)
   -h\!\left(\tfrac{1-s_2}{2}\right).
\label{eq:qkd-rate}
\end{equation}

\begin{figure}[t]
\centering
\includegraphics{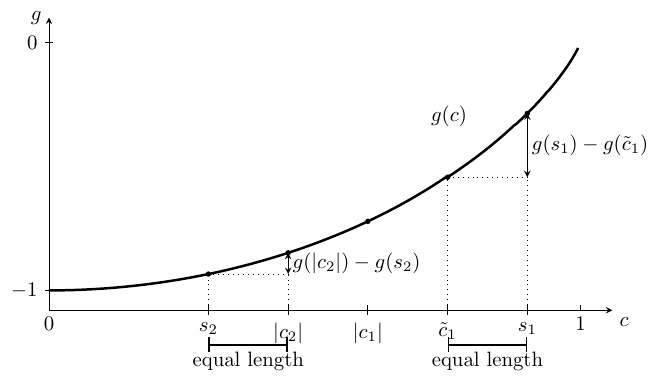}
\caption{The function $g(c)=-h((1-c)/2)$ on $0\le c\le1$, which is increasing
and convex.  The two marked intervals on the axis, $[s_2,|c_2|]$ and
$[\tilde c_1,s_1]$, have the same length, and the second lies to the right of
the first.  Because the slope of the curve increases to the right, the rise of $g$ over the
right interval, $g(s_1)-g(\tilde c_1)$, exceeds the rise $g(|c_2|)-g(s_2)$
over the left one.  The point $|c_1|$ lies to the left of $\tilde c_1$, so
$g(|c_1|)\le g(\tilde c_1)$ because $g$ is increasing, and it is enough to
bound $g(\tilde c_1)+g(|c_2|)$.}
\label{fig:gconvex}
\end{figure}

To prove Eq.~\eqref{eq:qkd-rate}, take the singular value decomposition
of $T_\rho$.  Its entries are expectation values of Hermitian operators and so
real, and the singular value decomposition writes it as an orthogonal matrix times a diagonal
matrix times the transpose of an orthogonal matrix,
$$
T_\rho=\begin{pmatrix}p_1&p_2&p_3\end{pmatrix}
\operatorname{diag}(s_1,s_2,s_3)
\begin{pmatrix}q_1&q_2&q_3\end{pmatrix}^{\mathsf T},
\qquad s_1\ge s_2\ge s_3\ge0 .
$$
The columns $p_1,p_2,p_3$ of the first factor and $q_1,q_2,q_3$ of the last
are orthonormal bases of $\mathbb R^3$.  Multiplying out the three factors
gives
$$
T_\rho=\sum_{k=1}^{3}s_k\,p_kq_k^{\mathsf T}.
$$
Let $\hat a_1,\dots,\hat a_m$ and $\hat b_1,\dots,\hat b_m$ be orthonormal vectors in
$\mathbb R^3$, so $m\le3$.
The singular value decomposition gives
$$
\sum_{i=1}^{m}\hat a_i^{\mathsf T}T_\rho \hat b_i=\sum_{k=1}^{3}s_kw_k,
\qquad
w_k=\sum_{i=1}^{m}(\hat a_i^{\mathsf T}p_k)(q_k^{\mathsf T}\hat b_i).
$$
Because $\hat a_1,\dots,\hat a_m$ are orthonormal, $\sum_i(\hat a_i^{\mathsf T}p_k)^2\le1$
for each $k$ and $\sum_k\sum_i(\hat a_i^{\mathsf T}p_k)^2=m$, and the same holds
for $\hat b_1,\dots,\hat b_m$.  The Cauchy--Schwarz inequality then gives $|w_k|\le1$ and
$\sum_k|w_k|\le m$.  A linear function of $w$ with coefficients
$s_1\ge s_2\ge s_3\ge0$ is largest under those two constraints when
$w_1=\dots=w_m=1$ and the remaining $w_k$ vanish, so
$\bigl|\sum_{i=1}^{m}\hat a_i^{\mathsf T}T_\rho \hat b_i\bigr|\le s_1+\dots+s_m$, with
equality at $\hat a_i=p_i$ and $\hat b_i=q_i$.  For $m=1$, since $a_i$ and $b_i$ are unit
vectors, $|c_i|\le s_1$ for each $i$.  For $m=2$, replacing $a_i$ by $-a_i$
where needed makes both $c_i$ nonnegative and leaves $a_1,a_2$ orthonormal, so
$|c_1|+|c_2|\le s_1+s_2$.  These two bounds say that $(|c_1|,|c_2|)$ is
weakly majorized by $(s_1,s_2)$.  The function
$g(c)=-h((1-c)/2)$ is increasing and convex on $0\le c\le1$, because $h$ is
concave and increasing on $0\le e\le\tfrac12$ and $(1-c)/2$ decreases from
$\tfrac12$ to $0$ as $c$ runs from $0$ to $1$.  For an increasing convex
function, weak majorization gives
$$
g(|c_1|)+g(|c_2|)\le g(s_1)+g(s_2).
$$
To see this, relabel the two basis pairs so that $|c_1|\ge|c_2|$.  If
$|c_2|\le s_2$, then $|c_1|\le s_1$ and $|c_2|\le s_2$ both hold, and the
inequality follows because $g$ is increasing.  Otherwise,
$s_2<|c_2|\le|c_1|\le s_1$, and $\tilde c_1=s_1+s_2-|c_2|$ satisfies
$|c_1|\le\tilde c_1\le s_1$, so $g(|c_1|)\le g(\tilde c_1)$.  The intervals
$[s_2,|c_2|]$ and $[\tilde c_1,s_1]$ have the same length, and the second lies
to the right of the first.  A convex function increases more over an interval
of given length the further to the right that interval lies
(Fig.~\ref{fig:gconvex}), so
$g(|c_2|)-g(s_2)\le g(s_1)-g(\tilde c_1)$, which rearranges to
$g(\tilde c_1)+g(|c_2|)\le g(s_1)+g(s_2)$.

The bracket in Eq.~\eqref{eq:qkd-rate} is $1+g(|c_1|)+g(|c_2|)$, so the
inequality $g(|c_1|)+g(|c_2|)\le g(s_1)+g(s_2)$
says that $r_{\rm opt}\le1+g(s_1)+g(s_2)$, the right-hand side of
Eq.~\eqref{eq:qkd-rate}.  That value is reached.  The choice $a_i=p_i$ and
$b_i=q_i$ for $i=1,2$ is admissible, since $p_1\perp p_2$ and $q_1\perp q_2$,
and it gives $c_i=p_i^{\mathsf T}T_\rho q_i=s_i$.

On the Werner line, the threshold has a closed form.  For
$\rho_\lambda(I/4)$,
$T_{\rho_\lambda}=\lambda\,\operatorname{diag}(1,-1,1)$, so
$s_1=s_2=\lambda$.  The $Z$ and $X$ bases therefore give
$e_b=e_p=e=(1-\lambda)/2$, and Eq.~\eqref{eq:qkd-rate} reduces to
$r_{\rm opt}=1-2h(e)$.  This bound vanishes when $h(e^*)=1/2$.  The
root in $[0,1/2]$ is $e^*=0.110028\ldots$, so the bound gives a positive key
only while the outcomes disagree in fewer than $11\%$ of the rounds, the
error tolerance of one-way BB84 with CSS codes~\cite{ShorPreskill2000}.
Translating back gives
$$
\lambda_{\rm BB84}=1-2e^*=0.779944\ldots\approx0.780 ,
$$
the weight of the horizontal guide in Fig.~\ref{fig:alignment-prl} above which
a Werner state gives a positive rate.
If $r_{\rm opt}\le0$, this bound certifies no key, but it does not rule
out key extraction through a stronger security analysis or a different
postprocessing protocol.

It remains to verify the hypotheses of Theorem~\ref{thm:phase} for the
set $\{\rho:r_{\rm opt}(\rho)\le0\}$.  Fix the two basis pairs.  Since
$h(e)=h(1-e)$, the absolute values in Eq.~\eqref{eq:qkd-rate} can be dropped,
and the bracket there is $1-h((1-c_1)/2)-h((1-c_2)/2)$, where each
$c_i=a_i^{\mathsf T}T_\rho b_i$ is linear in $\rho$.  Because $h$ is concave,
this bracket is a convex function of $\rho$.  The rate $r_{\rm opt}$ is its
maximum over the basis pairs, and a maximum of convex functions is convex, so
the set $\{\rho:r_{\rm opt}(\rho)\le0\}$ is convex.  By
Eq.~\eqref{eq:qkd-rate}, $r_{\rm opt}$ depends on $\rho$ only through the two
largest singular values of $T_\rho$.  A local unitary rotates each Bloch
sphere, which multiplies $T_\rho$ on the left and on the right by rotation
matrices and leaves its singular values unchanged.  Thus, $r_{\rm opt}$ is
invariant under local unitaries, and the set is closed under them.  The set
is also closed in the topological sense, since $r_{\rm opt}$ is continuous
in $\rho$, so its ability-absence interval is closed and, when nonempty,
contains the definitive threshold.

\subsubsection*{Noise that already has the ability}

The proof of Theorem~\ref{thm:phase} in the main text transfers membership
in $K$ one Bell weight at a time: if $\rho_\lambda(\sigma_X(\theta))\in K$,
then $\rho_\lambda(\sigma_X(\theta'))\in K$ for $\theta'\ge\theta$.  It
therefore gives the inclusion of ability-absence intervals
$I_K(\sigma_X(\theta))\subseteq I_K(\sigma_X(\theta'))$ without assuming that
either interval contains $\lambda=0$, and the threshold inequality
$\lambda_K(\sigma_X(\theta))\le\lambda_K(\sigma_X(\theta'))$ follows by taking
the supremum of each side.  Nothing in this requires the noise to lack the
ability, so noise that already has it is covered by the same proof.  As a
minimal example, take the entangled pure noise
$|\Phi_\theta\rangle=(|00\rangle+e^{-i\theta}|11\rangle)/\sqrt2$, whose
even-sector coherence is $u=\tfrac12e^{i\theta}$, and mix it
with $\PhiP$.  The separable interval is empty for every
$\theta<\pi$ and is exactly $\{\lambda=\tfrac12\}$ at $\theta=\pi$,
where the line passes through the even-sector classical mixture.
Thus, the definitive entanglement threshold rises from $0$ to
$\tfrac12$.  The inclusion
$\emptyset\subseteq\{\tfrac12\}$ is the widening in its simplest form.
The same example shows that no local unitary carries the mixing line at
some $\theta<\pi$ onto the one at $\pi$, weight by weight.  If one did, then
$K$ being closed under local unitaries would give the two lines the same
ability-absence interval and the same definitive threshold, whereas their
entanglement thresholds are $0$ and $\tfrac12$.

\subsubsection*{No local unitary removes the relative phase}

The example above covers only the pairs that end at $\theta=\pi$.  The
claim of the main text is the general one, over every $X$ noise state
and every pair of phases, and one invariant settles it.  A local unitary
rotates the two Bloch frames, so it sends the correlation tensor $T$ of a state
to $O_ATO_B^{\mathsf T}$, where $O_A$ and $O_B$ are rotations, and leaves
$\Tr(T^{\mathsf T}T)$ unchanged.  Write $w_\lambda=(1-\lambda)|v|$.  In the
local phase gauge that makes both coherences of $\sigma_X$ real and
nonnegative, the mixture $\rho_\lambda(\sigma_X(\theta))$ has
\begin{equation}
\Tr(T^{\mathsf T}T)=8q_\lambda(\theta)^{2}+8w_\lambda^{2}
+\bigl[\lambda+(1-\lambda)(a-b-c+d)\bigr]^{2}.
\label{eq:lu-invariant}
\end{equation}
The right-hand side of Eq.~\eqref{eq:lu-invariant} is strictly increasing in
$q_\lambda(\theta)^{2}$, and by the expression for $q_\lambda(\theta)^{2}$ in
the main text that quantity is strictly decreasing on $[0,\pi]$ whenever
$\lambda(1-\lambda)|u|>0$.  Two distinct phases in $[0,\pi]$ therefore give the
invariant two different values at every weight $0<\lambda<1$, so no local
unitary carries either mixing line onto the other.  The three excluded cases
are degenerate: at $\lambda=1$ the state is $\PhiP$ whatever $\theta$ is, at
$|u|=0$ the noise carries no coherence and so no phase, and at $\lambda=0$ the
state is the noise alone, whose phase the gauge does turn freely.

\refstepcounter{smsection}
\subsection*{S\arabic{smsection}. Figure reproducibility and numerical consistency checks}
\label{sec:sm-numerics}

This section gives the analytic curves and the certified $A\to B$
projective-measurement steerability intervals of Fig.~\ref{fig:alignment-prl},
together with the consistency checks behind its caption.  None of the
numerical checks below is used in the proof of Theorem~\ref{thm:phase} or
Theorem~\ref{thm:shadow}.
\subsubsection*{Analytic curves}

Figure~\ref{fig:alignment-prl}
uses the populations $(a,b,c,d)=(0.30,0.25,0.25,0.20)$ together with
$|u|=|v|=0.95\min\{\sqrt{ad},\sqrt{bc}\}=0.232702$, taking $u=|u|e^{i\theta}$
and $v=|v|$ in the local phase gauge.  These parameters define the family of the
relative-phase scan of Ref.~\cite{OperationalLayersCompanion}.  The entanglement, CJWR, and CHSH curves reproduce that scan, and the
BB84 curve and the certified steerability intervals are new.  Write
$a_\lambda=\tfrac{\lambda}{2}+(1-\lambda)a$, $b_\lambda=(1-\lambda)b$,
$c_\lambda=(1-\lambda)c$, $d_\lambda=\tfrac{\lambda}{2}+(1-\lambda)d$, and
$w_\lambda=(1-\lambda)|v|$ for the entries of the mixture.  The singular
values of its correlation tensor are
$$
\bigl\{\,2(q_\lambda+w_\lambda),\quad 2|q_\lambda-w_\lambda|,\quad
|a_\lambda-b_\lambda-c_\lambda+d_\lambda|\,\bigr\} ,
$$
which depend on $\theta$ only through $q_\lambda(\theta)$.  The BB84
curve is Eq.~\eqref{eq:qkd-rate} evaluated on the two largest of them.  The
entanglement and teleportation usefulness curves coincide because the
populations have $b=c$ and the noise state is separable.
The coincidence is a property of these populations rather than a general identity.

\textit{The anti-aligned endpoint meets the horizontal guides.}  At
$\theta=\pi$, using $a-b-c+d=0$ and $|u|=|v|$, the singular values are
$\lambda$, $\lambda$, and $|\lambda-4(1-\lambda)|u||$.  The last is at most
$\lambda$ for $\lambda\ge2|u|/(1+2|u|)=0.3176$, so there $s_1=s_2=\lambda$.
The anti-aligned CHSH-nonlocal threshold is therefore $1/\sqrt2$, from
$M=2\lambda^2$, and the anti-aligned BB84 threshold is $\lambda_{\rm BB84}$,
from $r_{\rm opt}=1-2h((1-\lambda)/2)$.  These are two of the three
horizontal guides of Fig.~\ref{fig:alignment-prl}.

\textit{The two caption claims at $\lambda=\tfrac12$.}  The caption of
Fig.~\ref{fig:alignment-prl} states that at $\lambda=\tfrac12$ the aligned
mixture gives a positive rate and the anti-aligned one is certified
unsteerable.  For the first, the aligned family has
$(s_1,s_2)=(0.965403,0.5)$ and $r_{\rm opt}=0.062731>0$, whereas the
anti-aligned family has $(s_1,s_2)=(0.5,0.5)$ and $r_{\rm opt}=-0.622556$.
For the second, the certified-unsteerable endpoint
at $\theta=\pi$ is $\lambda^{-}=\nngLamMinusPi>\tfrac12$ (Table~\ref{tab:sm-nng}), so the
anti-aligned mixture at $\lambda=\tfrac12$ admits a local-hidden-state model for all
projective measurements from $A$ to $B$.

\subsubsection*{Certified steerability intervals}

The $A\to B$ projective-measurement steerability threshold of the family
in Fig.~\ref{fig:alignment-prl}, written
$\lambda^{A\to B}_{\rm steer}$, is not known analytically.  We bracket it
with the critical-radius criterion of Ref.~\cite{NguyenNguyenGuhne2019}.  That
reference attaches to each two-qubit state a number $R$ and proves that the
state is steerable from $A$ to $B$ with projective measurements if and only if
$R<1$ [Eq.~(8) there].  The criterion is exact, but $R$ is not a closed-form
expression in the entries of the state: it is the value of the optimization in
Eqs.~\eqref{eq:sm-nng-radius} and~\eqref{eq:sm-nng-max} below.  The noise states of the
family are separable, since the partial transpose of an $X$ state is
positive exactly when $|u|^{2}\le bc$ and $|v|^{2}\le ad$, which the family
satisfies at every phase with $|u|=|v|=0.95\min\{\sqrt{ad},\sqrt{bc}\}$, so they are
unsteerable, and by convexity of the unsteerable set, the unsteerable weights
form the interval $[0,\lambda^{A\to B}_{\rm steer}]$.  Hence, $R\ge1$ up to the threshold and
$R<1$ above it.  Reference~\cite{NguyenNguyenGuhne2019}
also gives a way to bound $R$ from both sides, and we apply that method here to
compute a lower bound $R^{-}\le R$
and an upper bound $R^{+}\ge R$: a weight with $R^{-}\ge1$ lies below the
threshold and one with $R^{+}<1$ lies above it.

\emph{Canonical form.}  Reference~\cite{NguyenNguyenGuhne2019} states
the criterion for states in a normal form, reached as follows.  A two-qubit
state $\rho$ is fixed by its two marginal Bloch vectors, $\mathbf A$ for party $A$ and
$\mathbf B$ for party $B$, and by its correlation tensor $T$.  Their entries are
$A_j=\langle\sigma_j\otimes\openone\rangle$,
$B_k=\langle\openone\otimes\sigma_k\rangle$, and
$T_{jk}=\langle\sigma_j\otimes\sigma_k\rangle$.  A local filter on $B$ sends $\rho$
to $(\openone\otimes F)\rho(\openone\otimes F^{\dagger})$, up to
normalization, where $F$ is a matrix on the qubit of $B$.  A local filter on
$B$ maps an unsteerable state to an unsteerable state, so an unsteerable $\rho$
gives an unsteerable filtered state.  When $F$ is invertible, $F^{-1}$ is a
local filter that maps the filtered state back to $\rho$, so an unsteerable
filtered state gives an unsteerable $\rho$.  A steerable $\rho$ therefore
cannot give an unsteerable filtered state, since $F^{-1}$ would carry that
state back to $\rho$ and make $\rho$ unsteerable.  We call an inversion of one party's Bloch sphere the map that
sends every Bloch vector $\mathbf n$ of that party to $-\mathbf n$.  Applied to $B$, it takes
$(\mathbf A,\mathbf B,T)$ to $(\mathbf A,-\mathbf B,-T)$, and applied to $A$,
it takes them to $(-\mathbf A,\mathbf B,-T)$.  The critical radius $R$
is invariant under local unitaries on $A$, under invertible local filters on $B$
[Eq.~(11) of Ref.~\cite{NguyenNguyenGuhne2019}], and under inversion of either
Bloch sphere [Ref.~\cite{NguyenNguyenGuhne2019}, p.~3].
For the mixture of Eq.~\eqref{eq:line}, the marginal of $B$ is
$\rho_B=\lambda\openone/2+(1-\lambda)\sigma_B$, where $\sigma_B$ is the
marginal of the noise state.  The marginal $\rho_B$ has full rank at every $\lambda>0$, so
$\rho_B^{-1/2}$ exists.  Filtering by it takes the marginal of $B$ to
$\rho_B^{-1/2}\rho_B\rho_B^{-1/2}=\openone$, and renormalizing makes that
marginal $\openone/2$, so $\mathbf B=0$.  Write $T'$ for the correlation
tensor of the filtered state, and let $U\Sigma V^{\mathsf T}$ be its singular
value decomposition, where $U$ and $V$ are orthogonal and
$\Sigma=\operatorname{diag}(s_1,s_2,s_3)$ carries the singular values.  The
rotations of one party's Bloch vectors are exactly the maps that local
unitaries on that party's qubit produce.  Every orthogonal map on those Bloch
vectors is a rotation or an inversion followed by a rotation, so local
unitaries and inversions realize all of them.  Under orthogonal maps $O_A$ and
$O_B$ on the Bloch vectors of $A$ and $B$, the tensor $T'$ becomes
$O_AT'O_B^{\mathsf T}$, so
$O_A=U^{\mathsf T}$ and $O_B=V^{\mathsf T}$ give $\Sigma$, and finally, we
have

\begin{equation}
\mathbf B=0,\qquad T=\operatorname{diag}(s_1,s_2,s_3) .
\label{eq:sm-nng-canonical}
\end{equation}
To compute $R$ for the mixture of Eq.~\eqref{eq:line}, we then need
only the vector $\mathbf A$ and the singular values of $T$.  Reference~\cite{NguyenNguyenGuhne2019}
requires $T$ to be nonsingular, and it shows that a state with singular $T$ is
separable, hence unsteerable, so that case needs no criterion.

\emph{What we compute.}  Reference~\cite{NguyenNguyenGuhne2019} builds the
critical radius from a Hermitian operator $C$ on the qubit of $B$ and a
probability measure $\mu$ on the pure states $\omega$ of $B$.  Write
$\bar\rho=\rho-\tfrac12\openone\otimes\rho_B$.  We define the
Hilbert--Schmidt norm $\lVert X\rVert=\sqrt{\operatorname{Tr}(X^{\dagger}X)}$ for
an operator, and $\lVert\mathbf x\rVert$ for the length of a vector
of $\mathbb R^3$.  With the symbols of
this paper, Eq.~(7) there is

\begin{equation}
r(\rho,\mu)=\min_{C}\ \frac{1}{\sqrt2\,
\lVert\operatorname{Tr}_B[\bar\rho\,(\openone\otimes C)]\rVert}
\int d\mu(\omega)\,\lvert\operatorname{Tr}(C\omega)\rvert .
\label{eq:sm-nng-source}
\end{equation}
Equation~(5) there restricts $\mu$ to average to $\rho_B$.  The normal
form makes each piece of this explicit.  Every $C$ has the form
$C=\xi_0\openone+\sum_j\xi_j\sigma_j$, where $\xi_0$ is a number and
$\boldsymbol\xi$ is a vector of $\mathbb R^3$, so the minimum over $C$ is a
minimum over the nonzero vectors $(\xi_0,\boldsymbol\xi)$ of $\mathbb R^4$.  The pure state with Bloch vector
$\mathbf n$ is $\omega=\tfrac12(\openone+\sum_jn_j\sigma_j)$, so
$\operatorname{Tr}(C\omega)=\xi_0+\boldsymbol\xi\cdot\mathbf n$.  Pure states of $B$ and unit
vectors are in one-to-one correspondence, so $\mu$ carries over to a measure
$\nu$ on the unit sphere.  Averaging $\omega$ over $\nu$ gives
$\tfrac12(\openone+\sum_j\sigma_j\int n_j\,d\nu(\mathbf n))$, and the normal
form makes $\rho_B=\openone/2$, so that condition becomes
$\int\mathbf n\,d\nu(\mathbf n)=0$.  In the normal form,
$\bar\rho=\tfrac14(\sum_jA_j\,\sigma_j\otimes\openone
+\sum_{jk}T_{jk}\,\sigma_j\otimes\sigma_k)$.  With
$\operatorname{Tr}C=2\xi_0$ and $\operatorname{Tr}(\sigma_kC)=2\xi_k$,

\begin{equation}
\operatorname{Tr}_B[\bar\rho\,(\openone\otimes C)]
=\tfrac12\sum_j(\xi_0\mathbf A+T\boldsymbol\xi)_j\sigma_j .
\label{eq:sm-nng-trace}
\end{equation}
Its norm is $\lVert\xi_0\mathbf A+T\boldsymbol\xi\rVert/\sqrt2$ by
$\operatorname{Tr}(\sigma_j\sigma_k)=2\delta_{jk}$.  Writing $R(\nu)$ for $r(\rho,\mu)$,
Eq.~\eqref{eq:sm-nng-source} becomes

\begin{equation}
R(\nu)=\inf_{(\xi_0,\boldsymbol\xi)\neq0}\
\frac{\int d\nu(\mathbf n)\,\lvert \xi_0+\boldsymbol\xi\cdot\mathbf n\rvert}
{\lVert \xi_0\mathbf A+T\boldsymbol\xi\rVert} ,
\label{eq:sm-nng-radius}
\end{equation}
and the critical radius is the largest of these numbers [Eq.~(8)
there],

\begin{equation}
R=\max_{\nu}\,R(\nu),\qquad
\int\mathbf n\,d\nu(\mathbf n)=0 .
\label{eq:sm-nng-max}
\end{equation}
A measure on the whole sphere has infinitely many degrees of freedom,
so we approximate the sphere by a polytope with vertices
$\mathbf r_1,\dots,\mathbf r_n$, where $-\mathbf r_k$ is a vertex whenever
$\mathbf r_k$ is.  A
measure on the polytope is a list of weights $p_1,\dots,p_n$, and the
conditions on it are $p_k\ge0$, $\sum_kp_k=1$, and $\sum_kp_k\mathbf r_k=0$.
For such a measure, Ref.~\cite{NguyenNguyenGuhne2019} shows that the infimum in
Eq.~\eqref{eq:sm-nng-radius} is attained on a finite set of pairs, fixed by the
polytope and not by the measure, and that there are $O(n^3)$ of them
[Ref.~\cite{NguyenNguyenGuhne2019}, p.~4].  That set consists of the pairs for
which $\xi_0+\boldsymbol\xi\cdot\mathbf r=0$ at three of the vertices
$\mathbf r$.  We name such a pair by its plane, the points $\mathbf x$ with
$\xi_0+\boldsymbol\xi\cdot\mathbf x=0$, which passes through those three
vertices.
Reference~\cite{NguyenNguyenGuhne2019} gives no formula for such a pair, so we
compute it as follows.
Requiring $\xi_0+\boldsymbol\xi\cdot\mathbf x$ to vanish at $\mathbf r_i$, $\mathbf r_j$, and $\mathbf r_k$, and
subtracting the first equation from the other two, gives
$\boldsymbol\xi\cdot(\mathbf r_j-\mathbf r_i)
=\boldsymbol\xi\cdot(\mathbf r_k-\mathbf r_i)=0$.  Three distinct points of a
sphere are never collinear, so these two differences are independent, and a
vector of $\mathbb R^3$ orthogonal to both is a multiple of their cross
product.  We therefore take
$\boldsymbol\xi^{(\ell)}=(\mathbf r_j-\mathbf r_i)\times(\mathbf r_k-\mathbf
r_i)$, and then $\xi_0^{(\ell)}=-\boldsymbol\xi^{(\ell)}\cdot\mathbf r_i$,
which is the same number as $-\boldsymbol\xi^{(\ell)}\cdot\mathbf r_j$ and
$-\boldsymbol\xi^{(\ell)}\cdot\mathbf r_k$.  The
scale is immaterial because the numerator and the denominator of the ratio in
Eq.~\eqref{eq:sm-nng-radius} are homogeneous of degree one in the pair.  Different triples of vertices can lie on one plane,
and we keep one pair for each plane.
With the pairs indexed by $\ell$, the result of
Ref.~\cite{NguyenNguyenGuhne2019} above turns the infimum into a minimum, with
no approximation,

\begin{equation}
R(\nu)=\min_{\ell}\
\frac{\sum_kp_k\lvert\xi_0^{(\ell)}+\boldsymbol\xi^{(\ell)}\cdot\mathbf r_k\rvert}
{\lVert\xi_0^{(\ell)}\mathbf A+T\boldsymbol\xi^{(\ell)}\rVert} .
\label{eq:sm-nng-min}
\end{equation}
Clearing the denominator turns $R(\nu)\ge t$ into
$\sum_kp_k\lvert\xi_0^{(\ell)}+\boldsymbol\xi^{(\ell)}\cdot\mathbf r_k\rvert
\ge t\,\lVert\xi_0^{(\ell)}\mathbf A+T\boldsymbol\xi^{(\ell)}\rVert$ for every
$\ell$.  Each of these is linear in $p$ and $t$, because the two coefficients
are numbers fixed by the polytope and the state, so
maximizing $t$ is a linear program
[Ref.~\cite{NguyenNguyenGuhne2019}, p.~4], with
the $n$ weights and $t$ as unknowns and one constraint
for each plane,

\begin{align}
\max_{p,\,t}\quad & t\notag\\
\text{subject to}\quad & \textstyle\sum_k p_k\lvert \xi_0^{(\ell)}
  +\boldsymbol\xi^{(\ell)}\cdot\mathbf r_k\rvert
  \ \ge\ t\,\lVert \xi_0^{(\ell)}\mathbf A+T\boldsymbol\xi^{(\ell)}\rVert\quad\forall\ell,\notag\\
& \textstyle\sum_k p_k=1,\qquad \sum_k p_k\mathbf r_k=0,\qquad p_k\ge0 .
\label{eq:sm-nng-lp}
\end{align}

\emph{The two bounds.}  Put the vertices on the sphere, so that the
polytope lies inside it, and write $\varrho_{\rm in}$ for the radius of the
largest ball inside the polytope.  Every measure on this polytope is one of the
measures over which Eq.~\eqref{eq:sm-nng-max} takes its maximum, and a maximum
over fewer measures cannot be larger, so the value of
Eq.~\eqref{eq:sm-nng-lp} on this polytope is a lower bound, $R^{-}\le R$.  The polytope contains the ball of radius
$\varrho_{\rm in}$, so the same vertices scaled by $1/\varrho_{\rm in}$ give a
polytope containing the unit ball.  Every point of a polytope is a convex
combination of its vertices, so each unit vector is
$\mathbf n=\sum_kc_k(\mathbf n)\,\mathbf v_k$ over the scaled vertices
$\mathbf v_k=\mathbf r_k/\varrho_{\rm in}$, with $c_k\ge0$ and $\sum_kc_k=1$.  Put
$p_k=\int c_k(\mathbf n)\,d\nu(\mathbf n)$.  Then $p_k\ge0$, $\sum_kp_k=1$, and
$\sum_kp_k\mathbf v_k=\int\mathbf n\,d\nu(\mathbf n)=0$, so $p$ is a measure on
the scaled polytope with vanishing mean.  For every pair,

\begin{align}
\sum_kp_k\lvert\xi_0+\boldsymbol\xi\cdot\mathbf v_k\rvert
&=\int d\nu(\mathbf n)\sum_kc_k(\mathbf n)
   \lvert\xi_0+\boldsymbol\xi\cdot\mathbf v_k\rvert\notag\\
&\ge\int d\nu(\mathbf n)\Big\lvert\sum_kc_k(\mathbf n)
   (\xi_0+\boldsymbol\xi\cdot\mathbf v_k)\Big\rvert\notag\\
&=\int d\nu(\mathbf n)\,\lvert\xi_0+\boldsymbol\xi\cdot\mathbf n\rvert ,
\label{eq:sm-nng-outer}
\end{align}
by the definition of $p_k$, then the triangle inequality, then
$\sum_kc_k=1$ and $\sum_kc_k\mathbf v_k=\mathbf n$.  The denominator does not
involve the measure, so at every pair the ratio for $p$ is at least the ratio
for $\nu$, and the minimum over $\ell$ for $p$ is at least $R(\nu)$.  The
program of Eq.~\eqref{eq:sm-nng-lp} takes the largest value over all measures
on the scaled polytope, so it is at least $R(\nu)$, for every $\nu$.  Since $R$ is the largest $R(\nu)$, this
gives $R^{+}\ge R$ [Ref.~\cite{NguyenNguyenGuhne2019}, p.~4].
The outer vertices are $\mathbf r_k/\varrho_{\rm in}$, so the planes
through three of them are the pairs
$(\xi_0/\varrho_{\rm in},\boldsymbol\xi)$ built from the inner ones, and the
weights and the mean condition $\sum_kp_k\mathbf r_k=0$ are unchanged.  At such
a pair,
$\lvert\xi_0/\varrho_{\rm in}+\boldsymbol\xi\cdot(\mathbf r_k/\varrho_{\rm in})
\rvert=\lvert\xi_0+\boldsymbol\xi\cdot\mathbf r_k\rvert/\varrho_{\rm in}$, so
the numerator of Eq.~\eqref{eq:sm-nng-min} is $1/\varrho_{\rm in}$ times the
inner one, while the denominator
$\lVert(\xi_0/\varrho_{\rm in})\mathbf A+T\boldsymbol\xi\rVert$ is the inner
denominator with $\mathbf A$ replaced by $\mathbf A/\varrho_{\rm in}$.  Hence,
$R^{+}(\mathbf A,T)=R^{-}(\mathbf A/\varrho_{\rm in},T)/\varrho_{\rm in}$.
Ref.~\cite{NguyenNguyenGuhne2019} shows that $R^{-}$ does not increase when
$\mathbf A$ is scaled up (Theorem~31 of its appendix), so
$R^{+}(\mathbf A,T)\le R^{-}(\mathbf A,T)/\varrho_{\rm in}$.  Hence

\begin{equation}
R^{-}\le R\le R^{+},\qquad
\frac{R^{+}-R^{-}}{R^{-}}\ \le\ \frac{1}{\varrho_{\rm in}}-1 ,
\label{eq:sm-nng-sandwich}
\end{equation}
and the relative gap is fixed by the polytope alone.  We use $n=252$
vertices: for $i=0,\dots,125$,

\begin{align}
\mathbf r_i&=\big(\sqrt{1-z_i^{2}}\cos\phi_i,\
  \sqrt{1-z_i^{2}}\sin\phi_i,\ z_i\big),\notag\\
z_i&=\frac{i+1/2}{126},\qquad \phi_i=i\,\pi(3-\sqrt5) ,
\label{eq:sm-nng-fib}
\end{align}
together with $-\mathbf r_i$ for each $i$.  This set gives
$\varrho_{\rm in}=0.9817$ and $2\,611\,857$ planes.

\emph{The interval.}  For each weight $\lambda$, we take the canonical form of the state
$\rho_\lambda(\sigma_X(\theta))$ of Eq.~\eqref{eq:line}, which supplies
$\mathbf A$ and $T$, and solve the two programs for $R^{-}$ and $R^{+}$.  A
weight at which $R^{-}\ge1$ is unsteerable, and one at which $R^{+}<1$ is
steerable.  Neither $R^{-}<1$ nor $R^{+}\ge1$ decides anything, so a weight
with $R^{-}<1\le R^{+}$ is undecided and we run one search for each bound.
Write $\lambda^{-}$ for the largest weight we certify unsteerable, and
$\lambda^{+}$ for the smallest we certify steerable.
To find $\lambda^{-}$, we start from any two weights, one with $R^{-}\ge1$ and one
with $R^{-}<1$, and evaluate $R^{-}$ at the midpoint between them.  If
$R^{-}\ge1$ there, the midpoint replaces the first weight, and if not, it
replaces the second.  Each step halves the gap.  We stop once the gap is below $10^{-3}$, and report
the endpoint at which $R^{-}\ge1$, which is a lower bound on the steerability
threshold.  The
same search on $R^{+}$ gives $\lambda^{+}$, so

\begin{equation}
\lambda^{-}\le\lambda^{A\to B}_{\rm steer}\le\lambda^{+} .
\label{eq:sm-nng-interval}
\end{equation}
We run the search first on a $60$-vertex polytope, which is much
faster and gives rough values of $\lambda^{-}$ and $\lambda^{+}$.  We then
repeat the search on the $252$-vertex polytope, which is finer and brackets
more tightly, and we report only its values.  There are thirty-two reported
endpoints, two for each of the sixteen phases, and the smallest singular value
of $T$ among them is $0.131$, so $T$ is nonsingular throughout.

Between two sampled phases, the bracket still holds, and that is the shaded
band of Fig.~\ref{fig:alignment-prl}.  Write $\theta_k$ and $\theta_{k+1}$ for
consecutive samples.  Theorem~\ref{thm:phase} makes
$\lambda^{A\to B}_{\rm steer}$ nondecreasing in $\theta$, so for
$\theta\in[\theta_k,\theta_{k+1}]$ it lies between
$\lambda^{A\to B}_{\rm steer}(\theta_k)$ and
$\lambda^{A\to B}_{\rm steer}(\theta_{k+1})$, and Eq.~\eqref{eq:sm-nng-interval}
at the two ends gives
$\lambda^{-}(\theta_k)\le\lambda^{A\to B}_{\rm steer}(\theta)\le
\lambda^{+}(\theta_{k+1})$.  The band is that step envelope, so nothing is
interpolated between the sampled phases.

\emph{Implementation.}  We solve the linear programs in Python with
NumPy and SciPy.  A program has one constraint for each plane, too many to
hand the solver at once.  We solve it with a small subset.  That solution
need not satisfy the constraints we left out.  We test
it against every plane, add the ones it fails, the worst first, and repeat
until it fails none, at which point it solves the full program.
The point we stop at is feasible for all the constraints and optimal for a
program with fewer of them, so it is optimal for the full program.
\emph{Numerical accuracy.}  Everything is evaluated in double precision,
and we claim no exact arithmetic.  Rounding in the sums over the vertices
in Eq.~\eqref{eq:sm-nng-min} enters near $10^{-13}$, while the quantities that
decide an endpoint, $R^{-}-1$ at $\lambda^{-}$ and $1-R^{+}$ at $\lambda^{+}$,
are at least $6\times10^{-7}$ at the thirty-two endpoints of
Table~\ref{tab:sm-nng}.  Those endpoints are printed to $10^{-4}$, which is enough
to indicate the nondecreasing trend in $\theta$.  The monotonicity of the
threshold in Theorem~\ref{thm:phase} is proved analytically and does not rest
on these numbers.

\begin{table}[t]
\caption{Certified $A\to B$ projective-measurement steerability interval
$[\lambda^{-},\lambda^{+}]$ at the sampled phases, from
Eq.~\eqref{eq:sm-nng-interval} with $n=252$ and $\varrho_{\rm in}=0.9817$.
Each endpoint is rounded away from the interval, $\lambda^{-}$ downward and
$\lambda^{+}$ upward, so the printed interval contains the certified one.  For
example, at $\theta=0$ the certified $0.137593$ and $0.164193$ are printed as
$0.1375$ and $0.1642$.}
\label{tab:sm-nng}

\renewcommand{\arraystretch}{1.15}
\begin{tabular}{cccc}
\hline\hline
$\theta/\pi$ & $\lambda^{-}$ & $\lambda^{+}$ & width \\
\hline
$0$ & $0.1375$ & $0.1642$ & $0.0267$ \\
$1/15$ & $0.1420$ & $0.1687$ & $0.0267$ \\
$2/15$ & $0.1553$ & $0.1826$ & $0.0273$ \\
$1/5$ & $0.1793$ & $0.2076$ & $0.0283$ \\
$4/15$ & $0.2174$ & $0.2444$ & $0.0270$ \\
$1/3$ & $0.2665$ & $0.2917$ & $0.0252$ \\
$2/5$ & $0.3232$ & $0.3457$ & $0.0225$ \\
$7/15$ & $0.3823$ & $0.4009$ & $0.0186$ \\
$8/15$ & $0.4360$ & $0.4526$ & $0.0166$ \\
$3/5$ & $0.4829$ & $0.4963$ & $0.0134$ \\
$2/3$ & $0.5207$ & $0.5333$ & $0.0126$ \\
$11/15$ & $0.5505$ & $0.5618$ & $0.0113$ \\
$4/5$ & $0.5734$ & $0.5832$ & $0.0098$ \\
$13/15$ & $0.5890$ & $0.5982$ & $0.0092$ \\
$14/15$ & $0.5977$ & $0.6067$ & $0.0090$ \\
$1$ & $0.6011$ & $0.6096$ & $0.0085$ \\
\hline\hline
\end{tabular}
\end{table}

\end{document}